\documentclass[superscriptaddress,twocolumn,pra]{revtex4-1}

\usepackage{amsfonts}
\usepackage{amsmath}    
\usepackage[normalem]{ulem}
\usepackage{bm}
\usepackage{graphicx}
\usepackage{titlesec}
\usepackage{tabularx}
\usepackage{color}
\usepackage{bbold}
\usepackage{upgreek,ulem,textcomp}
\usepackage{bbold}
\usepackage{comment}
\usepackage[british]{babel}
\usepackage{csquotes}
\usepackage{etoolbox}

\usepackage[colorlinks=true,linkcolor = black,urlcolor=blue, citecolor = black]{hyperref}
\usepackage{hypernat}

\newcommand{\ket}[1]{\vert#1\rangle}
\newcommand{\bra}[1]{\langle#1\vert}

\newcommand{\RNum}[1]{\uppercase\expandafter{\romannumeral #1\relax}} 

\begin{abstract}
Electromagnetically induced transparency (EIT) and Autler-Townes splitting (ATS) are similar, but different quantum optical phenomena: EIT results from a Fano interference, whereas ATS is described by the AC-Stark effect. Likewise, despite their close resemblance, light-storage techniques based on the EIT memory protocol and the recently-proposed ATS memory protocol (E. Saglamyurek \emph{et al.} \textit{Nature Photonics} \textbf{12}, 2018) are distinct: the EIT protocol relies on adiabatic elimination of absorption, whereas the ATS protocol is based on absorption. In this article, we elaborate on the distinction between EIT and ATS memory protocols through numerical analysis and experimental demonstrations in a cold rubidium ensemble. We find that their storage characteristics manifest opposite limits of the light-matter interaction due to their inherent adiabatic vs.\ non-adiabatic nature. Furthermore, we determine optimal memory conditions for each protocol and analyze ambiguous regimes in the case of broadband storage, where non-optimal memory implementations can possess characteristics of both EIT and ATS protocols. We anticipate that this investigation will lead to deeper understanding and improved technical development of quantum memories, while clarifying distinctions between the EIT and ATS protocols.  
\end{abstract}

\begin{document}

\title{Discerning quantum memories based on electromagnetically-induced-transparency and Autler-Townes-splitting protocols}

\author{Anindya Rastogi}
\thanks{These authors contributed equally to the work.}
\affiliation{Department of Physics, University of Alberta, Edmonton AB T6G 2E1, Canada}
\author{Erhan Saglamyurek}
\thanks{These authors contributed equally to the work.}
\affiliation{Department of Physics, University of Alberta, Edmonton AB T6G 2E1, Canada}
\author{Taras Hrushevskyi}
\affiliation{Department of Physics, University of Alberta, Edmonton AB T6G 2E1, Canada}
\author{Scott Hubele}
\affiliation{Department of Physics, University of Alberta, Edmonton AB T6G 2E1, Canada}
\author{Lindsay J. LeBlanc}
\affiliation{Department of Physics, University of Alberta, Edmonton AB T6G 2E1, Canada}
\affiliation{Canadian Institute for Advanced Research, Toronto, ON, Canada}
\email{Correspondence to: lindsay.leblanc@ualberta.ca}

\maketitle

\section{INTRODUCTION} \label{sec:intro}

Interactions between electromagnetic fields and a three-level atomic system provide a wealth of opportunities for probing many quantum optical phenomena.  Among those, the Autler-Townes effect~\cite{Autler1955}, first demonstrated more than sixty years ago, results in a ``split'' transition within a coupled three-level system, due to the AC-Stark effect. The Autler-Townes splitting (ATS) emerges in the strong-coupling limit of the more-recently discovered electromagnetically induced transparency (EIT)~\cite{Boller1991}. The EIT effect is described by the formation of a dark-state due to a destructive quantum interference between transition pathways~\cite{Fleischhauer2005}. Both ATS and EIT result in a transparency feature, which is qualitatively identified as a wide spectral region between the split-absorption peaks for ATS, but a narrow transmission window within a single absorption peak for EIT. This common feature has been at the centre of a long-standing confusion as to whether an observed transparency is due to EIT or ATS, and as such, the distinction between the two is an active topic of research in the quantum optics community. Recently, it has been shown theoretically~\cite{Abi-Salloum2010,Anisimov2011a,He2015} as well as experimentally~\cite{Giner2013,Sun2014,Lu2015,Tan2014,Hao2017} that it is possible to objectively distinguish the regime where EIT dominates (EIT regime) from the one where ATS dominates (ATS regime). 

Despite these important results, their direct connection to applications remains largely unexplored. Two most significant applications of EIT, the slow-light effect~\cite{VestergaardHau1999,Kash1999,Phillips2001,Liu2001,Turukhin2002,Longdell2005,Safavi-Naeini2011} and the related quantum memory approach (EIT memory protocol)~\cite{Eisaman2005a,Lvovsky2009,Heshami2016} have been extensively studied. However, ATS has not garnered nearly as much attention~\cite{Sparkes2010,Liao2014}. This oversight may be partly due to a common misconception that quantum interference~ is a necessary feature for coherent storage, and partly due to lack of interest in the ATS regime where the technical demand would be very high for a slow-light based quantum memory. 

In this context, the recently proposed and demonstrated ATS quantum memory protocol~\cite{Saglamyurek2017} has emerged as a direct application of the ATS effect. This protocol relies on controlled absorption of light through ATS peaks, in contrast to the EIT-protocol that is based on adiabatic elimination of absorption. Despite this fundamental distinction, these protocols may still be confused with each other, just as has been the case with the EIT and ATS phenomena. This is due to the fact that the EIT and ATS protocols bear close similarities and common technical features, which may give a wrong impression that the ATS protocol is simply the EIT protocol operating in the ATS regime. Furthermore, as explored in this study, the transition between the EIT and ATS memory protocols is smooth and can inadvertently happen by simply tuning the ``knob'' of the coupling field in the laboratory. This fact may easily lead to misinterpretations of light-storage implementations, particularly from an experimentalist's point of view. Finally, discerning between EIT and ATS memories is important not only to eliminate potential confusion, but also to develop a practical quantum memory by choosing the protocol best-suited to each set of technical and design limitations.

In this study, we make a detailed comparison between EIT and ATS protocols from both fundamental and technical perspectives.  Our numerical~ analyses conclusively show that the differences between the protocols are beyond those related to the physical regimes that are tied to the protocols' names. Essentially, their storage mechanisms exhibit contrasting aspects of the light-matter interaction, including in-phase vs.\ out-of-phase spin-photon dynamics, dispersion vs.\ absorption based signal delay, shape- vs.\ pulse-area-based control-field optimization, and adiabatic vs.\ non-adiabatic operation. We further analyze optimal memory conditions for the two protocols in the EIT and ATS regimes, showing that the ATS memory protocol is intrinsically suitable for broadband operation, whereas the EIT protocol is  well-suited for narrowband operation. This also implies that while ATS memory cannot be implemented in the EIT regime for narrowband signals, the implementation of an EIT memory in the ATS regime for broadband storage, although possible, is technically demanding compared to ATS memory. Additionally, we investigate ambiguous cases in which a broadband memory implementation can feature a ``mixed'' character of both EIT and ATS protocols, and quantitatively distinguish such cases from the ``true'' EIT and ATS memory operation. Finally, we complement our analysis with experimental demonstrations of the ATS and EIT protocols in an ensemble of laser-cooled Rb atoms. These experiments highlight the key differences between these protocols, and also demonstrate how varying experimental parameters leads to transition between them. 

We believe that our study will eliminate some of the existing, as well as potential, misconceptions in regards to EIT and ATS memories, while bringing an application-based perspective on the phenomenon-based comparisons between EIT and ATS regimes. This study also provides practical recipes for developing high-performance optical quantum memories using the EIT and ATS protocols under realistic conditions with limited resources.

\begin{figure}[t!]
\begin{center}
\includegraphics[width = 85mm] {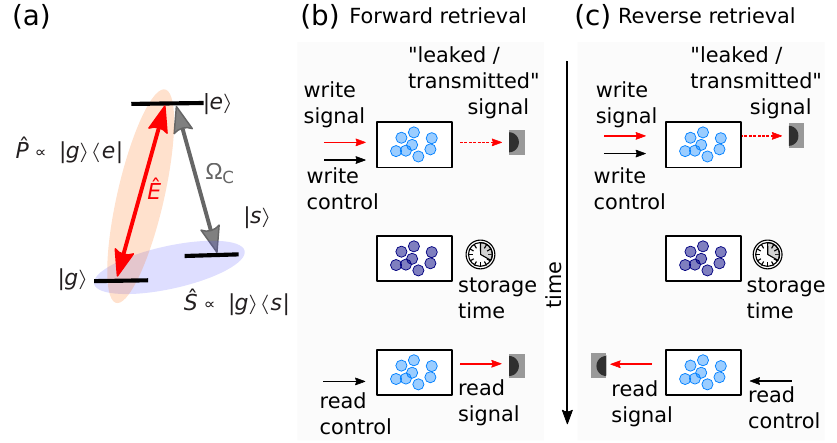}
\caption{(a) A three-level $\Lambda$-type atom or atom-like system, coupled by a weak signal field $E(z,t)$ between $\ket{g}$ and $\ket{e}$ levels and a strong control field with Rabi frequency $\Omega_{\rm C}$ between $\ket{s}$ and $\ket{e}$ levels. The atomic coherences are indicated as shaded regions with polarization coherence $\hat{P} \sim \ket{g}\bra{e} + h.c.$ (orange) and the spin coherence $\hat{S} \sim \ket{s}\bra{g} + h.c.$. (blue). (b) Forward retrieval scheme with co-propagating write and read control pulses. The output photonic mode is emitted in the propagation direction of the input mode. (c) Backward retrieval scheme with counter-propagating write and read control pulses. The output photonic mode is emitted at the input side of the medium.}
\label{fig:levels}
\end{center}
\end{figure}

\section{THEORETICAL ANALYSIS} \label{sec:Theory}
\subsection{Maxwell-Bloch description of EIT and ATS memory protocols} 
\label{sec:MBE}
For side-by-side theoretical comparisons between the EIT and ATS memory protocols, we employ the Maxwell-Bloch equations~\cite{Gorshkov2007a,Gorshkov2007b,Liao2014} to numerically analyze these schemes in an ensemble of $\Lambda$-type atomic systems with long-lived spin-based ground levels (Fig.~\ref{fig:levels}a). We assume that all atoms (atom-number $N$) are initially in the ground state $\ket{g}$, and an incoming weak ``signal'' field (to be delayed or stored) is resonant with $\ket{g} \leftrightarrow \ket{e}$ transition. A strong coupling  field (referred to as ``control'') with Rabi frequency $\Omega_{\rm C}$ (much larger than that of the signal) resonantly drives the $\ket{s} \leftrightarrow \ket{e}$ transition, where $\Omega_{\rm C} = \bra{e} \mathbf{d}\cdot \mathbf{E}_{\rm C} \ket{s}/\hbar$, $\mathbf{d}$ is the electric-dipole operator, and $\mathbf{E}_{\rm C}$ is the control electric field. The combined decoherence rate for $\ket{e} \leftrightarrow \{\ket{g},\ket{s}\}$ optical transitions is $\gamma_{\rm e} = \Gamma / 2$, and the decoherence rate between the ground levels is $\gamma_{\rm s}$. We assume that $\gamma_{\rm s} \ll \gamma_{\rm e}$. Under these conditions, the Maxwell-Bloch equations are, \begin{align}
&(\partial_t + c \partial_z) \hat{E}(z,t) =  i g \sqrt{N} \hat{P}(z,t),\label{eq:MB1}\\
&\partial_t \hat{P}(z,t) = - \gamma_{e} \hat{P}(z,t) \!+\! i g \sqrt{N} \hat{E}(z,t) + \frac{i}{2} \Omega_{\rm c} \hat{S}(z,t),\label{eq:MB2}\\
&\partial_t \hat{S}(z,t) = -\gamma_\textrm{\rm s} \hat{S}(z,t)+ \frac{i}{2} \Omega_{\rm c}^* \hat{ P}(z,t), \label{eq:MB3}
\end{align}
where $\hat{E}(z,t)$ is the electric field operator for the photonic field, and $\hat{P}(z,t)$  and $\hat{S}(z,t)$ are the polarization and spin-wave operators, that describe the collective atomic coherences of the $\ket{g}\leftrightarrow\ket{e}$ and $\ket{g}\leftrightarrow\ket{s}$ transitions, respectively~\cite{Gorshkov2007b}. The strength of the atom-light coupling is $g\sqrt{N}=\sqrt{{c d  \gamma_{e}}/{2L}}$, where $d$ is the peak optical depth and $L$ is the length of the atomic medium along $z$, which is the propagation direction of the photonic field.

In our numerical analysis, we focus on two kinds of coherent memory processes:~a fixed-delay process, where the output signal emerges at a predetermined time while the control field is held constant throughout; and an on-demand process, where the retrieval time of the output signal is controlled by a time-dependent control field $\Omega_{\rm C}(t)$. We consider input pulses with a Gaussian temporal profile characterised by a full-width half-maximum time $\tau$ and bandwidth $B$~[related as $B = (2 \ln2 / \pi)\times \tau$]. To evaluate the mechanisms and the efficiency of the storage-retrieval processes, we are interested in the output photonic mode given by the electric field, either at the exit [$\mathcal{E}_{\rm out} = \hat{E} (L,t)$: forward recall] or at the entrance [$\mathcal{E}_{\rm out} = \hat{E} (0,t)$: backward recall] of the storage medium, and the spin- and polarization-mode coherences $S(z,t)$ and $P(z,t)$ throughout the medium.

Our analysis for on-demand memory operation is based on the ``optimality'' criterion, which states that the maximally achievable efficiency is uniquely determined by the optical depth and is independent of the protocol used~\cite{Gorshkov2007a,Gorshkov2007b} (hereafter, such implementations are referred to as~``optimal''). For optimal EIT and ATS memories, we use a control-field optimization specific to that protocol. We note that, since optimality in general, requires complete time-reversal of the system dynamics, our analysis is based on the backward retrieval configuration for the output signal mode [Fig.~\ref{fig:levels}(c)], unless otherwise stated. 

We systematically compare EIT and ATS protocols in both narrow- and broadband signal regimes. The signal bandwidth is considered narrowband when $B < \Gamma/2\pi$ (equivalently $\tau > 1/\gamma$), and broadband when $B > \Gamma/2\pi$ ($\tau < 1/\gamma$). We note that there is a well-defined connection between the memory operation in these bandwidth regimes and the physical regimes described by EIT ($F \equiv \Omega_{\rm C} / \Gamma < 1$) and  ATS ($F > 1$) phenomena (Sec.~\ref{sec:disp vs abs}). In the broadband regime, optimal memory implementation for both EIT and ATS protocols falls into the ATS regime. In the narrowband regime, depending on the optical depth, optimal memory using the EIT protocol can be implemented in either ATS or EIT regime. On this basis, we emphasize that the optimal memory protocol -- EIT or ATS -- is not necessarily tied to the physical operating regime of the same name.  As such, the analysis that follows compares the operation of the EIT protocol in the narrow- vs.\ broadband regimes, and the EIT vs.\ ATS protocols in the broadband regime. Comparisons involving narrowband ATS memory are not explicitly shown in most of the presented results, since the ATS scheme cannot operate in the native EIT regime with any reasonable efficiency~(Sec.~\ref{sec:adiabaticity}).  

\begin{figure*}[tb!]
\begin{center}
\includegraphics[width = 17.0 cm] {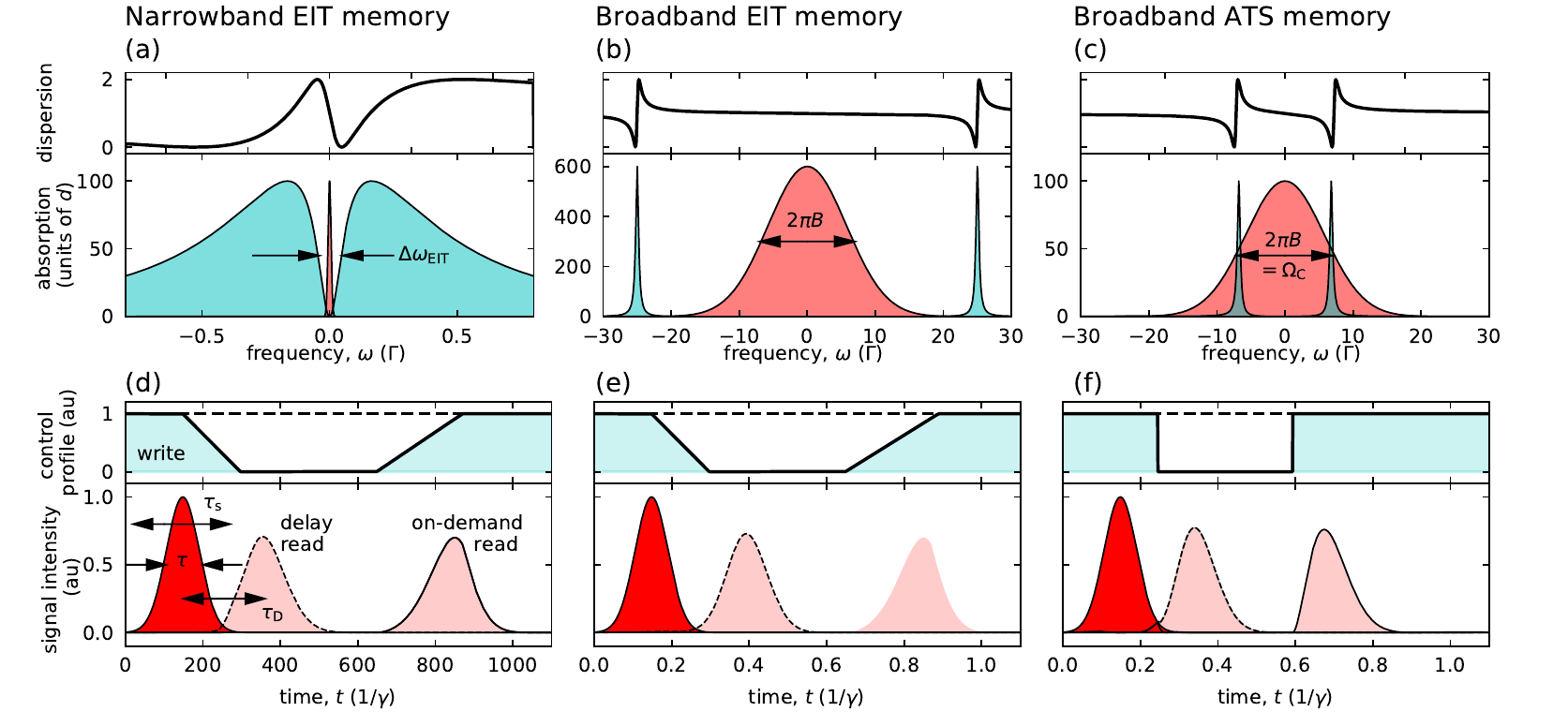}
\caption{\textbf{(a-c)} Spectral domain representation of signal delay, for the parameters given in (d-f). The signal bandwidth (red) relative to the medium's absorption profile (blue) determines the dominant mechanism that would induce delay (Sec.~\ref{sec:disp vs abs}). Absorption and input profiles are given in units of optical density and the associated dispersion curves (black) are shown in the upper panel. \textbf{(d-f)} Memory implementations via the EIT and ATS protocols for narrow- and broadband signals. A Gaussian input probe (dark red) can be subjected to a predetermined delay (light red and dashed curve) under a constant control field; or can be stored and subsequently retrieved after a desired storage time (light red and solid curve) via an interrupted control field (black). In each case, the parameters are adjusted to achieve a memory efficiency of $\eta = 90\%$, corresponding to the optimal efficiency at $d=100$. (d) Narrowband EIT protocol with $B=0.014 \Gamma/2\pi$, $d=100$, and $\Omega_{\rm C}^{\rm pk}=0.5 \Gamma$. (e) Broadband EIT protocol with $B=14 \Gamma/2\pi$, $d=600$, and $\Omega_{\rm C}^{\rm pk}=55\Gamma$. (f) Broadband ATS protocol with $B=14 \Gamma/2\pi$, $d=100$, and $\Omega_{\rm C}^{\rm pk}=14\Gamma$. The animations depicting the coherence dynamics in the storage and on-demand recall processes for the above configurations are available in supplementary information.}
\label{fig:timeFreq}
\end{center}    
\end{figure*}

\subsection{Comparing EIT and ATS memory protocols} \label{sec:Numerical Results}
The EIT and ATS memory schemes resemble each other in many ways: both use a relatively strong, resonant control field to store and retrieve a weak, resonant signal field. Both can be described by the same set of Maxwell-Bloch equations when operating under the same limits (e.g. $\Omega_{\rm C} > \Gamma$). Comparing their time-domain pictures [Fig.~\ref{fig:timeFreq}(d-f)] further suggests
11
 that both operate in almost the same way:~under a constant-control (time-invariant $\Omega_{\rm C}$) the input signal undergoes some delay through the atomic medium before being re-emitted (dashed curves in the figure). If the control is turned off before the signal leaves the medium, the signal is stored as a collective spin-excitation (writing stage). When the control field is turned back on after a desired time, the signal is retrieved on-demand (read-out stage). 

These similarities may give the impression that ATS and EIT memories are identical. However, the underlying physical mechanisms of storage and  retrieval are quite different, and clear distinctions emerge when comparing the dynamics of the spin and photonic coherences (Sec.~\ref{sec:coherence}), the nature of the delay mechanism (Sec.~\ref{sec:disp vs abs}), procedures for control field optimization (Sec.~\ref{sec:Control optimization}), and optimal memory conditions (Sec.~\ref{sec:adiabaticity}). 

\subsubsection{Coherence dynamics} \label{sec:coherence}
The reversible transfer of coherence between photonic and spin modes is a common feature in most on-demand memory schemes, including the EIT and ATS protocols. However, mechanisms underlying this mapping depend on the protocol. From this perspective, we compare the coherence dynamics of the EIT and ATS protocols and discuss their essential differences in both narrow- and broadband signal regimes. These differences lay the foundation for the remaining sections.

\begin{figure*}[tb!]
\begin{center}
\includegraphics[width = 17.0 cm] {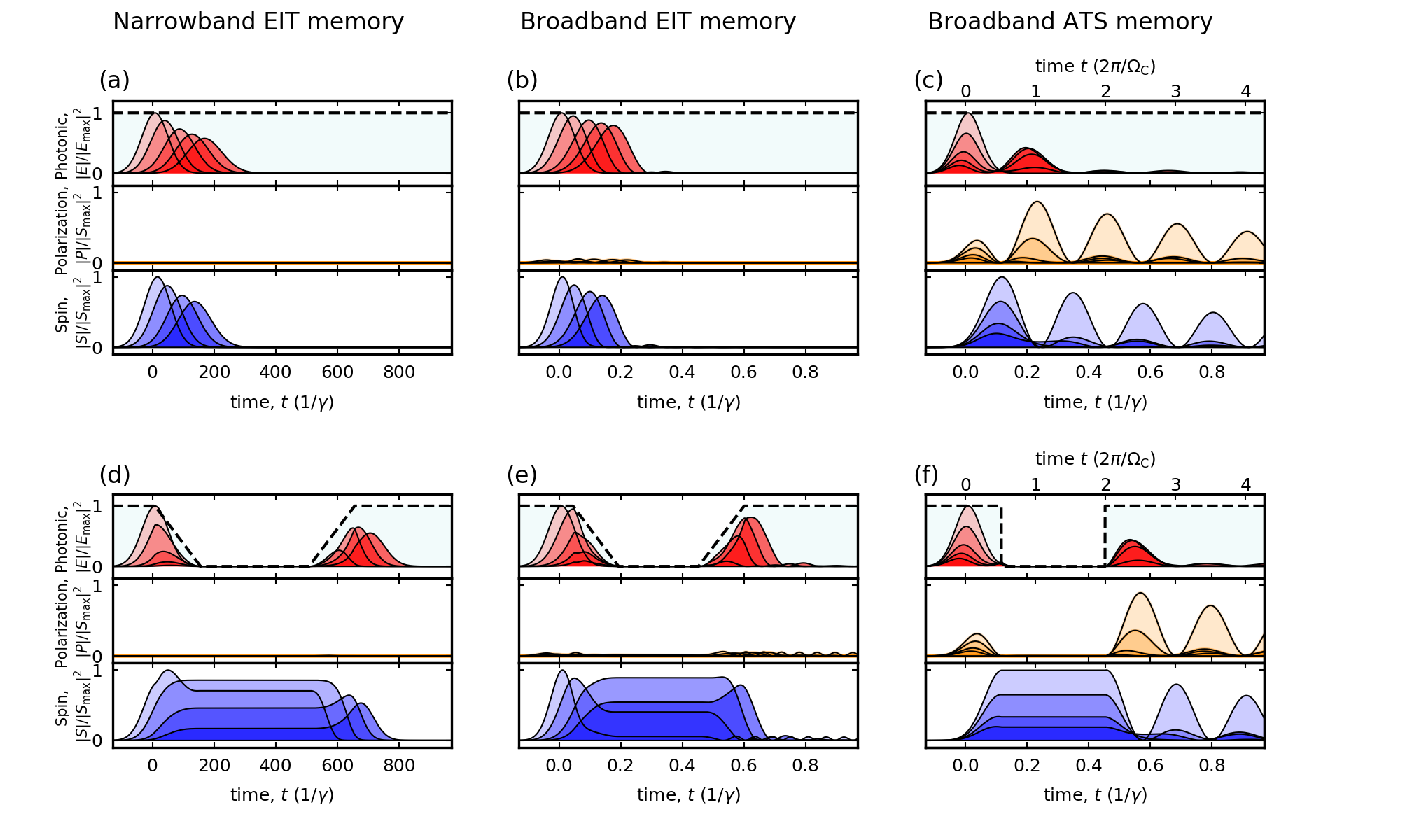} 
\caption[System Dynamics]{Temporal evolution of normalized coherences in the photonic $|E(z,t)|^2$ (red, upper), polarization $|P(z,t)|^2$ (orange, middle), and spin $|S(z,t)|^2$ (blue, lower) modes for the EIT protocol in the narrow-(a,d) and broadband (b,e) signal regimes, and for the ATS protocol in the broadband regime (c,f). The polarization coherence is normalized with respect to the maximum value of spin. The lightest shading shows coherence at the entrance to the medium ($z = 0$), with progressively darker shading towards later sections at $L/5, L/2$ and $3L/4$. The coherence dynamics are analyzed for  delay (a,b,c) and on-demand memory (d,e,f) operations in the forward retrieval scheme, such that the output photonic mode is emitted at $z = L$. The dashed line indicates the normalized $\Omega_{\rm C}$ profile. The parameters $\{\tau, B, d, \Omega_{\rm C}^{\rm pk}\}$ are: (a,d) $\{100/\gamma, 0.014\Gamma/2\pi$, ${\rm 40}$, $0.33\Gamma\}$; (b,e) $\{0.1/\gamma, 14\Gamma/2\pi, {\rm 800}, 55\Gamma\}$; (c,f) $\{0.1/\gamma, 14\Gamma/2\pi, {\rm 40}, 14\Gamma\}$. For the ATS protocol, the periodic exchange of coherence between $P$ and $S$ modes occurs at the control field Rabi frequency.} 
\label{fig:CoherenceEvolution}
\end{center}
\end{figure*}

Figure~\ref{fig:CoherenceEvolution} illustrates the evolution of the photonic, polarization, and spin coherences for the narrowband [Figs.~\ref{fig:CoherenceEvolution}(a,d)] and broadband  [Figs.~\ref{fig:CoherenceEvolution}(b,e)] EIT protocols as well as broadband ATS protocol [Figs.~\ref{fig:CoherenceEvolution}(c,f)]  for both the constant and interrupted (i.e. switched off and then back on) control fields. In the EIT protocol, when a constant control field is applied~[Figs.~\ref{fig:CoherenceEvolution}(a,b)], the signal is essentially transmitted through the medium that is normally opaque in the absence of the control. This is due to a destructive quantum interference between the two transition pathways~[$\ket{g} \leftrightarrow \ket{e}$ and $\ket{s} \rightarrow \ket{e}$ in Fig.~\ref{fig:levels}(a)], that eliminates the coupling of the signal to the excited level, thereby forming a dark-state.~Inside the EIT medium, some fraction of the photonic coherence is dynamically transferred to the spin-excitation while the group velocity of the remaining photonic component is substantially reduced, leading to a spatially compressed photonic mode [top panels in Fig.~\ref{fig:CoherenceEvolution}(a,b)]. Importantly, the spin and the associated compressed photonic mode propagate together [bottom panels of Fig.~\ref{fig:CoherenceEvolution}(a,b)], with the same group-velocity (orders of magnitude smaller than the vacuum speed of light) until the photonic mode exits the medium as a delayed pulse. This phenomenon is at the heart of the dark-state polariton description of the well-known slow-light effect \cite{Fleischhauer2000}~(alternatively described in the spectral domain representation as in Sec.~\ref{sec:disp vs abs}). In this picture, the polariton is a superposition of the photonic and spin modes with the associated probabilities:
\begin{align}
& |S(z,t)|^{\rm 2} = \frac{g^{\rm 2} N}{\big(g^{\rm 2} N + \Omega_{\rm C}^{\rm 2}(t)\big)} \label{eqn:eit_spin} \\
& |E(z,t)|^{\rm 2} = \frac{\Omega_{\rm C}^{\rm 2}(t)} {\big(g^{\rm 2} N + \Omega_{\rm C}^{\rm 2}(t)\big)}, \label{eqn:eit_photonic}
\end{align}
and, ideally, zero probability for the coherence to be in the polarization mode. 

Furthermore, as indicated by Eqs~(\ref{eqn:eit_spin}) and (\ref{eqn:eit_photonic}), it is possible to entirely stop the slowly moving photonic mode by adiabatically reducing $\Omega_{\rm C}(t)$ to zero, which allows for a complete transfer of coherence into the stationary spin-wave mode, effecting storage (writing), as shown in Figs.~\ref{fig:CoherenceEvolution}(d,e). After a desired time, switching the control back on remaps the stored coherence on to the photonic-mode leading to an on-demand retrieval of the signal (read-out). 

Comparing Figs.~\ref{fig:CoherenceEvolution}(a,b) and \ref{fig:CoherenceEvolution}(d,e) shows that although the basic principle of the EIT protocol is the same in both the narrow- and broadband regimes, there are two important differences: first, unwanted absorption manifests as an incoherent process that causes signal loss via spontaneous emission in the former, whereas it leads to coherent processes that contribute to storage in the latter. This effect plays a significant role in assessing the true ``EIT-memory-character'' in the broadband operation regime (Sec.~\ref{sec:Control optimization}). Second, the resources required to implement an optimal EIT memory differ considerably between narrow- and broadband regimes. As an example, for the parameters in Fig.~\ref{fig:CoherenceEvolution}, broadband EIT operation ($B = 14\Gamma/2\pi$)  requires 20 times larger optical depth than the narrowband EIT~($B = 0.014\Gamma/2\pi$). The bandwidth scaling of optical depth and control power are further explored in (Sec.~\ref{sec:Control optimization}-\ref{sec:adiabaticity}). 

The time-evolution of coherences in the ATS protocol is remarkably distinct from the EIT memory. First, the ATS protocol [Figs.~\ref{fig:CoherenceEvolution}(c,f)] relies on the transfer of coherence to the polarization mode, unlike in the EIT protocol where it is adiabatically  eliminated (middle panels in Fig.~\ref{fig:CoherenceEvolution}). Second, in contrast to the slow-light-based delay for the EIT protocol, signal delay in the ATS scheme is due to the periodic exchange of coherence between spin and photonic modes as mediated by the polarization mode \cite{Saglamyurek2017}:
\begin{align}
& |S(z,t)|^{\rm 2} \propto \cos^{\rm 2}(\Omega_{\rm C} t) \label{eqn:ats_spin} \\
& |P(z,t)|^{\rm 2} \propto \sin^{\rm 2}(\Omega_{\rm C} t) \label{eqn:ats_polarization} \\
& |E(z,t)|^{\rm 2} \propto \sin^{\rm 2}(\Omega_{\rm C} t). \label{eqn:ats_photonic}
\end{align}
Within these dynamics lies a key difference between protocols: unlike the in-phase dynamics of the EIT protocol [Eqs~(\ref{eqn:eit_spin})-(\ref{eqn:eit_photonic})], the spin- and photonic-coherence dynamics in the ATS protocol are in quadrature with each other [Eqs~(\ref{eqn:ats_spin})-(\ref{eqn:ats_photonic})].

Finally, on-demand storage and recall operations in the ATS memory protocol relies on pausing the process of coherence exchange after a time interval $t = 2\pi /\Omega_{\rm C}$, when the coherence is completely in the spin mode, by abruptly switching off the control-field [Fig.~\ref{fig:CoherenceEvolution}(f)], waiting for a storage time $T$ and switching the control-field back on [Fig.~\ref{fig:CoherenceEvolution}(f)]. This mechanism, thus, differs from the EIT protocol [Figs.~\ref{fig:CoherenceEvolution}(d,e)] where storage is achieved by adiabatically decoupling the spin mode from the photonic mode by gradually ramping off the control-field, and the retrieval is stimulated by the reverse process.  

\begin{figure}[tb!]
\begin{center}
\includegraphics[width = 7.5 cm] {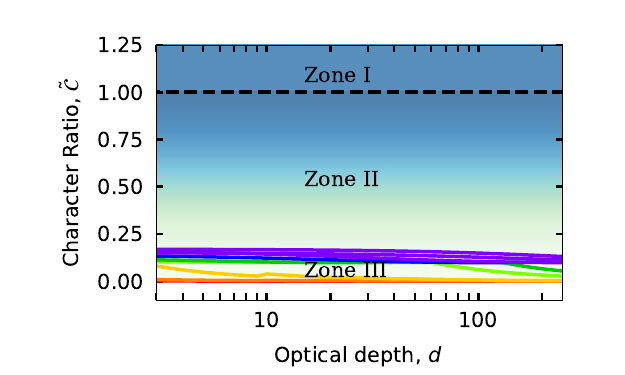}
\caption[Memory-character factor]{Normalized memory-character factor $\tilde{\mathcal{C}}$ as a function of optical depth for various signal bandwidths, extending from the narrow~($B = 0.0007\Gamma/2\pi$) to the broadband regime ($B = 136\Gamma/2\pi$) for the EIT protocol. Pure-EIT and -ATS storage correspond to $\tilde{\mathcal{C}} \lesssim 0.1$ (Zone~\RNum{3}) and $\tilde{\mathcal{C}} \gtrsim 1$ (Zone~\RNum{1}), respectively. The progression from light to dark shading represents transition from EIT to ATS storage. For narrow bandwidths (red, orange, yellow), the values of $\tilde{\mathcal{C}}$ stay well below the threshold.
For improperly optimized EIT protocols, the memory operation lies within Zone~\RNum{2}, indicating mixed storage due to the presence of both EIT and ATS mechanisms. Optimal ATS operation is shown by $\tilde{\mathcal{C}}=1$ (dashed line).} \label{fig:character_ratio}
\end{center}    
\end{figure}

Though the numerical analysis reveals starkly different physical mechanisms underlying EIT- and ATS memories, from an experimental standpoint, the control field is the single knob that leads to these distinctions. Particularly in the broadband signal regime, both EIT and ATS protocols can be implemented efficiently (albeit with different optimality conditions) by engineering the control field according to the protocol of interest. In cases where the control field is not properly optimized for either protocol, the storage and recall processes can exhibit the character of both ATS and EIT protocols at the same time (Sec.~\ref{sec:Control optimization}). To distinguish such cases from the standard EIT and ATS protocols, we define a dimensionless parameter $\mathcal{C}$ that quantifies the character of a memory:
\begin{align}
\mathcal{C} = \frac{1}{\tau_{\rm s}}\frac{\int_{\rm 0}^{\rm L}\int_{\rm 0}^{\rm \tau_{\rm s}}|P(z,t)|^{\rm 2} dz~ dt}{\int_{\rm 0}^{\rm L}|S(z,T)|^{\rm 2}dz}.
\end{align}
This parameter gives the ratio of the average normalized polarization coherence during the writing period ($0< t < \tau_{\rm s}$) to the normalized spin coherence measured at some point after writing is complete and the signal is stored ($T> \tau_{\rm s}$). Here and elsewhere, we define the writing period as $\tau_{\rm s} = 2.25\tau$. The memory character $\mathcal{C}$ can be used to determine whether a memory is characterized by ``pure-EIT'' protocol, a ``pure-ATS'' protocol, or a mix of the EIT/ATS protocols. In Fig.~\ref{fig:character_ratio}, we show that a control-field-optimized ATS memory yields a constant value of $\mathcal{C} \equiv \mathcal{C}_{\rm 0}$ independent of the optical depth and signal bandwidth, and this serves as a normalization factor in our comparisons. We also note that while the absolute value of $\mathcal{C}_{\rm 0}$ may show some variation with different control profiles,~it remains fixed for a given optimal control.

In contrast,~a control-field-optimized EIT memory has a $\mathcal{C}$ parameter at least an order of magnitude smaller than that of the optimized ATS. Therefore, in our analysis, we normalize all character ratios with respect to the $\mathcal{C}_{\rm 0}$ value by using $\tilde{\mathcal{C}} = \mathcal{C}/\mathcal{C}_{\rm 0}$.  Defining the limits:~$\tilde{\mathcal{C}} = 1$ and $\tilde{\mathcal{C}} = 0.1$ to represent the true ATS and EIT memory operations respectively, the intermediate values correspond to hybrid memory operation. 

Along similar lines, we can define a delay-character parameter $\mathcal{C}_{\rm D}$ to determine whether a signal delay is dominated by the EIT or ATS mechanism (Sec.~\ref{sec:disp vs abs}). This parameter quantifies the degree of polarization-coherence elimination over a chosen time interval (longer than the signal delay time), and thus helps in identifying the slow-light and the ATS character of a signal delay. However, while the general character of a memory is revealed by this parameter, the numerical limits are less well-defined than for $\mathcal{C}$.

\subsubsection{Dispersion-vs-absorption based signal delay} \label{sec:disp vs abs}
Under constant-control-field conditions, signal pulses are subject to a fixed delay, which is mediated by the slow-light effect for EIT-protocols and by absorption and reemission for ATS protocols. Whether a signal is delayed via the EIT or ATS protocol depends on the signal bandwidth relative to the absorption spectrum in the atomic medium, which is determined by the strength of the control field.~In this section, we analyze the delay characteristics of EIT and ATS protocols. Our emphasis is on configurations with sufficiently large delays ($\tau_{\rm d} \gtrsim \tau_{\rm s}$) that lead to optimal on-demand memory operations under realistic conditions (yielding efficiencies up to 95$\%$ with reasonable optical depths). 

For the delay of a narrowband signal via the EIT protocol [Fig.~\ref{fig:timeFreq}(d)], a control field in the limit $\Omega_{\rm C} \lesssim \Gamma$ (which is the EIT regime when $d \le 200$) induces a narrow transparency window of width $\Delta \omega_{\rm EIT} = {\Omega_{\rm c}^{\rm 2}}/{\sqrt{d}\Gamma}$~\cite{Fleischhauer2005,Novikova2012}, such that the signal bandwidth falls well within this window ($B < \Delta \omega_{\rm EIT}/2\pi$), as seen in Fig.~\ref{fig:timeFreq}(a). While this condition prevents signal absorption,~the steep linear dispersion in the vicinity of the narrow transparency window results in a group delay given by $\tau_{\rm D}^{\rm EIT} = {d \Gamma}/{\Omega_{\rm C}^2}$. This delay is usually expressed as fractional delay, given by $\tau_{\rm D}^{\rm EIT} / \tau$, which is a measure of the pulse-fraction that can be spatially trapped (compressed) inside the EIT medium.    

For the delay of a broadband signal via the EIT scheme~[Fig. \ref{fig:timeFreq}(e)], a wide-transparency window is necessary and can be obtained only with large values of control Rabi frequency, which eventually tends towards operation in the Autler-Townes regime ($\Omega_{\rm C} >\Gamma)$, where the absorption lines are split by spacing of $\delta_{\rm ATS} \approx \Omega_{\rm C}$. In this regime, the previous expression for the workable width of the transparency ($\Delta \omega_{\rm EIT}$) is no longer valid, and if applied, the ATS peaks would lie inside the transparency window: $\delta_{\rm ATS} < \Delta \omega_{\rm EIT}$. Instead, we find that absorption is substantially eliminated when $\delta_{\rm ATS}/2\pi$ is significantly larger than the bandwidth [typically, $\delta_{\rm ATS}/2\pi \gtrsim 4B$, as shown in Fig.~\ref{fig:timeFreq}(b)]. Under this condition, the shallow gradient of dispersion between the ATS peaks leads to sufficiently large fractional delays ($\tau_{\rm D}/\tau > 1$) only with very large optical depths.

In contrast, broadband signal delay via the ATS protocol [Fig.~\ref{fig:timeFreq}(f)] relies on signal absorption and hence, a significant spectral overlap is necessary between the signal bandwidth and the ATS peaks.This condition is optimally satisfied when $\delta_{\rm ATS}/2\pi = \Omega_{\rm C}/2\pi = B$ [Fig.~\ref{fig:timeFreq}(c)], and the resulting delay $\tau_{\rm D}^{\rm ATS} = 2\pi / \Omega_{\rm C}$ is due to the re-emission of light (following the signal absorption) after one Rabi period. In the ATS protocol, smaller and non-uniform dispersion negates the possibility of effective slow light, making the coherent absorption and re-emission processes dominant. Therefore, efficient broadband signal delay via the ATS protocol, together with sufficient delay ($\tau_{\rm D}/\tau > 1$), can be accomplished with, at least, an order of magnitude less optical depth than the corresponding EIT protocol.  

\begin{figure}[tb!]
    \centering
     \includegraphics[width = 8.5cm]{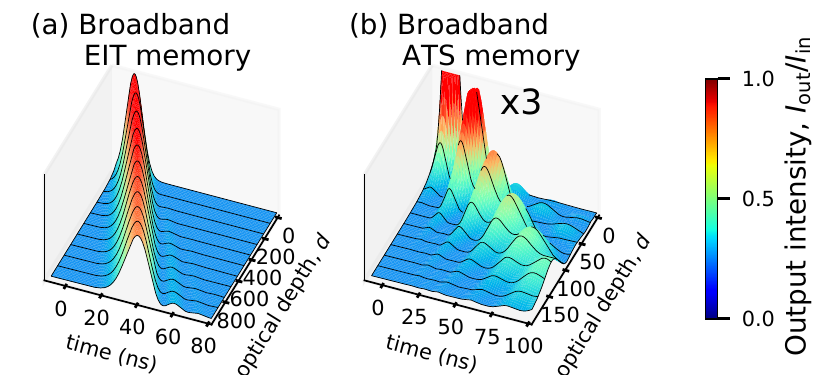}
    \caption[OD dependence of signal delay]{Optical-depth dependence of the delay characteristics in (a) EIT and (b) ATS protocols for a broadband signal with $B= 7\Gamma/2\pi$ under forward retrieval. Colorbar indicates the normalized intensity of delayed pulses. (a) A constant control with $\Omega_{\rm C} = 28\Gamma$ results in a slow-light-mediated group delay which monotonically increases with the optical-depth. Here, the signal delay is smooth and continuous unlike the discretized delays in the ATS protocol. (b) A constant control with $\Omega_{\rm C} = 2\pi B = 7\Gamma$ generates a fixed delay of one Rabi period $\tau_{\rm D} = 2\pi/\Omega_{\rm C} \approx 24 ns$ for all $d \lesssim 3F$. Large optical depths result in the suppression of first order ``echo'' (re-absorption) followed by higher order emissions occurring at integer multiples of the Rabi period.}
   \label{fig:surf}
\end{figure}

Directly comparing the optical-depth dependence of signal delay in the EIT and ATS protocols reveals important differences [Fig.~\ref{fig:surf}(a,b)]. In the EIT protocol, for a fixed bandwidth and $\Omega_{\rm C}/2\pi = 4B$, slow-light delay increases smoothly as optical depth is increased.  Equivalently, the delay-bandwidth-product for EIT protocol is $\tau_{\rm D}^{\rm EIT} \Delta \omega_{\rm EIT} = \sqrt{d}$.  In contrast, the ATS-protocol delay is constant and equal to the Rabi period for optical depths $d\lesssim 3F$, and results in a fixed delay-bandwidth product $\tau_{\rm D}^{\rm ATS} \Delta\omega_{\rm ATS} = 2\pi$. For larger optical-depths, ($d > 3F$), the re-emitted (delayed) pulse undergoes re-absorption and subsequent re-emission, resulting in multiple output pulses at integer multiples of the first delay-time~($\tau_{\rm D}^{\rm ATS}$). 

The multiple, higher-order emissions in the ATS protocol are reminiscent of the same behaviour found in photon-echo memory approaches, such as the well-studied atomic frequency comb (AFC).~The AFC protocol relies on controlled dephasing and rephasing of the atomic polarization through an imprinted comb-shaped spectral feature \cite{Afzelius2009, Afzelius2010}. Further comparisons between the ATS and the AFC protocol yield more insight: in the AFC protocol, a signal whose bandwidth spans at least two teeth (\emph{stationary} absorption peaks) of the comb is subject to a delay, determined by the inverse of the peak spacing.~The ATS delay exhibits the same character since the control Rabi frequency \emph{dynamically} determines the spacing between the ATS lines that are spanned by the signal spectrum. Moreover, as in the ATS protocol, while the delay of the AFC echo is independent of optical depth, relatively large effective optical-depths result in high-order echoes at integer multiples of the first AFC delay. However, if the signal bandwidth is reduced to a level where it lies well between the two peaks of the AFC, i.e., inside a ``spectral-hole'', the resulting delay is determined by the associated dispersion feature, which depends on the optical depth and the width of the hole itself, thereby, exhibiting the same character as the EIT-based slow-light effect.

It is worth pointing out that the common features between different delay mechanisms (for example, the ATS delay due to the polarization-mediated coherence exchange between spin and photonic modes, or the AFC delay due to periodic dephasing and rephasing of polarization) are a consequence of the fact that signal delay through a linear system (in these cases, the quantum memory medium) depends only on the shape of absorption (and associated dispersion) profile of that system regardless of the physical origin (i.e..~light-induced absorption peaks of ATS or stationary absorption combs of the AFC). This principle is the essence of linear spectral-filtering theory for classical signal processing and can also be used to describe the presented features of the different delay mechanisms~\cite{Kusunoki1998,Bonarota2012,Chaneliere2018} as an alternative to the Maxwell-Bloch treatment in this work.

\begin{figure}[tb!]
    \centering
    \includegraphics[width = 8.5 cm]{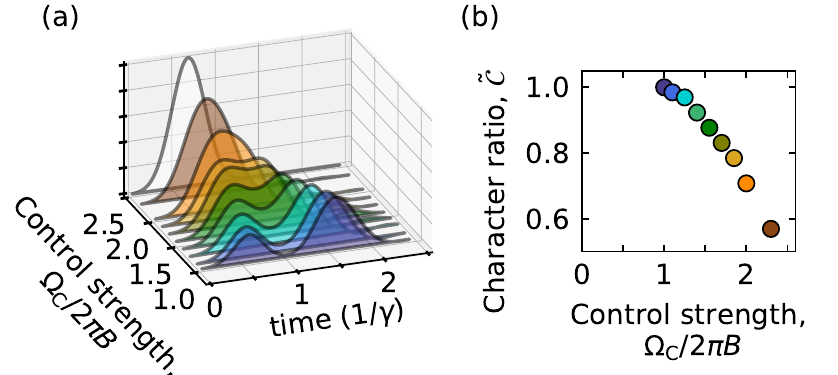}
    \caption{Mixed EIT-ATS delay in the intermediate signal regime by varying the control strength in the range $B \le \Omega_{\rm C}/2\pi < 2.5B$, where $B = 14\Gamma/2\pi$ and optical depth $d = 30$. (a) The ATS delay (blue), using the forward retrieval scheme, is shown as a reference. Upon increasing $\Omega_{\rm C}$, the distinct transmitted and delayed peaks tend to merge towards each other indicating a mixed storage character due to both EIT- and ATS-based delay mechanisms. For $\Omega_{\rm C}/2\pi \gtrsim 2B$, the two peaks are no longer separate and EIT delay starts to dominate. (b) The associated delay character values. Transition from ATS- to EIT-based delay is shown by the decrease in $\tilde{\mathcal{C_{\rm D}}}$-values.} 
    \label{fig:delay_character}
\end{figure}

Finally, we investigate the delay process in the intermediate regime \big($B < \Omega_{\rm C}/2\pi < 4B$\big), where the broadband signal does not satisfy either the ATS or the EIT condition.  Here, signal delay is analyzed with respect to the control Rabi frequency for a fixed bandwidth and optical depth  [Fig.~\ref{fig:delay_character}(a)], and the associated delay-character parameter is calculated for each setting~ [Fig.~\ref{fig:delay_character}(b)]. These analyses, together with the calculated $\mathcal{C}_{\rm D}$-values show that in the regimes of $\Omega_{\rm C}/2\pi = B$ and $\Omega_{\rm C}/2\pi = 4B$, the resulting delay is purely characterized by mechanisms based on ATS and EIT delays, respectively.  However, for the intermediate regime, the delay exhibits a dual character showing simultaneously, the basic features of the absorption/re-emission and slow-light based delays.

\subsubsection{Shape-based vs. pulse-area-based control-field optimization} \label{sec:Control optimization}
Control-field engineering, which varies the timing and strength of the field, plays a key role in optimizing EIT and ATS memories, but through different mechanisms. Moreover, we find that in the broadband regime, a non-optimized control (i.e., one that is not fully compatible with either protocol) can lead to storage and retrieval processes that exhibit a mixed character~(with intermediate values of $\tilde{\mathcal{C}}$), which may be efficient, but are still non-optimal.

In general, control-field optimization in the EIT protocol aims to minimise the signal absorption, whereas in the ATS protocol, it aims to maximize the same. For the EIT protocol in the narrowband regime, the control-field mediates trapping (spatial compression) of the signal via the slow-light effect, while adiabatically transferring coherence to the spin-wave mode. Insufficient spatial compression (associated with large $\Omega_{\rm C}$ values) means that the signal pulse does not fit inside the finite-length medium, and thus cannot be completely stored.~This loss is referred to as the ``leakage loss''. On the other hand, larger compression (associated with small $\Omega_{\rm C}$) and/or non-adiabatic coherence transfer leads to signal absorption, which results in loss through spontaneous emission (``absorption loss'').  The control-field optimization therefore, involves finding a trade-off between these two loss mechanisms by properly ``shaping'' the field in two degrees of freedom: the \emph{strength} and the \emph{temporal profile}. The strength optimization is typically achieved with a control intensity that results in fractional delay of $\tau_{\rm D}/\tau \approx 2$ \cite{Hsiao2018}, which also demands large optical depths (typically $d > 20$).~The temporal-profile optimization is relatively straightforward for such optical depths, requiring the switch off/on part to be sufficiently smooth to preserve adiabatic dynamics during the entire storage and retrieval processes. However, for small optical depths $(d < 20)$, it is not possible to achieve a fractional delay of 2 (without significant loss and distortion of the signal) by a simple adjustment of the control-field strength. In this situation, the profile must not only maintain the adiabatic evolution, but must also mediate the best possible signal compression with minimal transmission loss. Since such optimization depends strongly on optical depth and the profile of the input field, it is typically a non-trivial task~\cite{Gorshkov2007b,Novikova2007,Novikova2008,Phillips2008,Chen2013}. 

\begin{figure}[tb!]
    \centering
    \includegraphics[width = 8.5 cm]{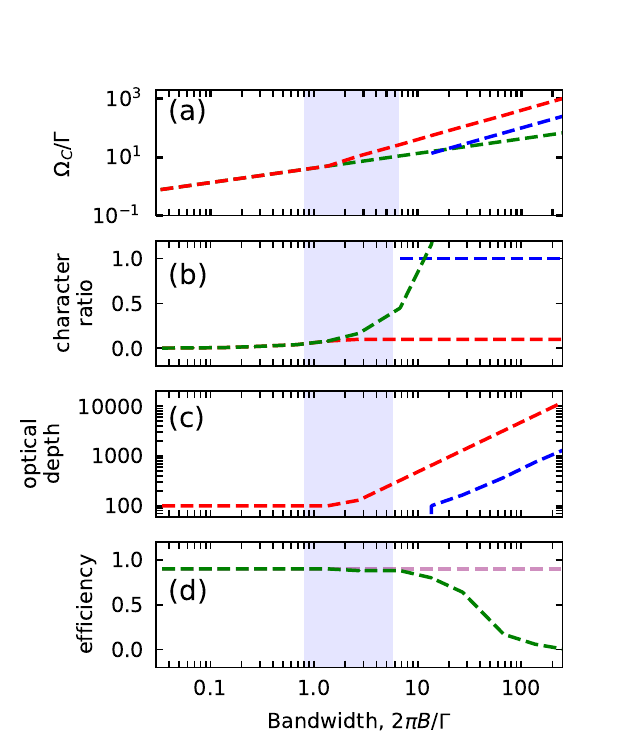}
    \caption[Scaling of resources with bandwidth]{Resource scaling with respect to the bandwidth for EIT (red), and ATS (blue) protocols along with the narrow-band EIT scaling (green) applied in the broadband regime. (a) Control power scaling, (b) Character ratio scaling,  (c) Optical depth scaling, and (d) Efficiency scaling.  For each bandwidth, $\Omega_{\rm C}$ and $d$ are adapted to maintain a memory efficiency $\eta = 90\%$.  For narrowband signals, the EIT protocol gives the optimal efficiency for $d = 100$ and $\Omega_{\rm C}^{\rm EIT} \propto \sqrt{B}$. Maintaining the same $\Omega_{\rm C}^{\rm EIT}$ scaling (with fixed $d$) for broadband signals results in a non-optimal ATS-based storage as shown by the dramatic rise in $\tilde{\mathcal{C}}$ and drop in the efficiency values (green). The shaded region, where $B \sim \Gamma/2\pi$, indicates the crossover between EIT and ATS phenomena, as well as between the protocols. For true broadband EIT memory, we set $\Omega_{\rm C}^{\rm EIT}/2\pi = 4B$ and scale $d$  such that $\tau_{\rm D}^{\rm EIT} / \tau \approx 2$, thereby maintaining $\tilde{\mathcal{C}}$ at its threshold limit. For the broadband ATS protocol, $\Omega_{\rm C}^{\rm ATS}/2\pi = B$ and the effective optical depth is fixed such that $d = 6F$.}
\label{fig:character}
\end{figure}

Although the general optimization strategy for the EIT protocol is the same for both narrow- and broadband signal regimes, there are two important distinctions in terms of the strength and temporal profile. First, since in these regimes the slow-light effect is mediated by either a narrow transparency feature or broadly separated ATS lines (Sec.~\ref{sec:disp vs abs}), the corresponding spectral conditions differ as $B < \Delta \omega_{\rm EIT}/2\pi$ and $B < \Omega_{\rm C}/8\pi$, respectively. When considering the general requirements needed to achieve an optimal EIT memory, this difference leads to different scalings of the control field strength with respect to signal bandwidth. In the narrowband regime, at a constant optical depth, any signal may be optimally stored by scaling the control strength as $\Omega_{\rm C} \propto \sqrt{B}$ (thereby keeping the fractional delay fixed),~as shown in Figs.~\ref{fig:character}(a-d).  As the bandwidth approaches the transition linewidth ($B\approx\Gamma/2\pi$), the $\mathcal{C}$ parameter tends to increase, but is still an order of magnitude smaller than for the optimal ATS scheme [Fig.~\ref{fig:character}(b)],~indicating the EIT-character of the storage. 
If the same square-root scaling of $\Omega_{\rm C}$ is applied in the broadband signal regime (Fig.~\ref{fig:character}a), the memory is predominantly characterised by the mechanism of the ATS protocol, although with non-optimal efficiency [Fig.~\ref{fig:character}(d)]. This occurs because the ATS peaks begin to spectrally overlap the signal bandwidth ($\delta_{\rm ATS}/2\pi \lesssim B$), as is also evident by the increase in $\mathcal{C}$-values [Fig.~\ref{fig:character}(a-b)].  Therefore, in order to maintain the EIT-memory character with optimal efficiency in this regime, the control Rabi frequency, together with optical depth, must be scaled linearly with $B$ (while at the same time, maintaining $\Omega_{\rm C}/2\pi \approx 4B$), as shown in Figs.~\ref{fig:character}(a-d). 

\begin{figure}
\centering
\includegraphics[width = 8.5 cm]{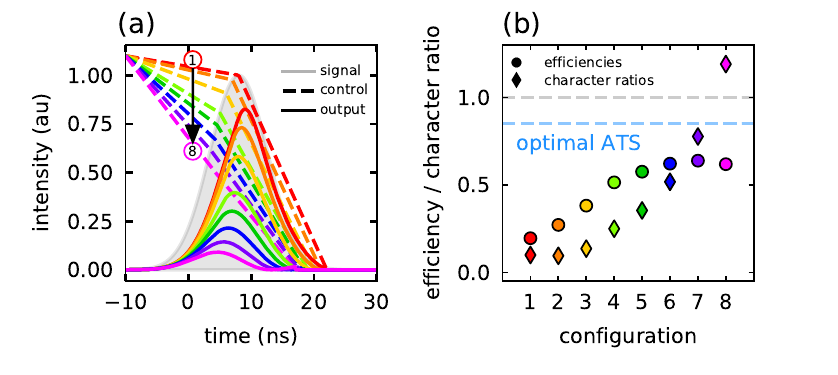}
\caption{Transition from the EIT to ATS storage mechanism by variation of the temporal-profile of the control-field in the broadband regime. Here, $B = 7\Gamma/2\pi$, $\Omega_{\rm C}^{\rm pk} = 28 \Gamma$, and $d = 60$. (a) Writing stage  control field profiles, characterized by different slopes during the switch-off (dashed).  Also shown are the Gaussian input (shaded gray) and the leaked (for EIT-dominated storage, as in configurations 1-3) or  transmitted (for ATS-dominated storage, as in configurations 4-8) pulses. ``Configuration 1'' corresponds to the true-EIT storage with $\tilde{\mathcal{C}} = 0.1$ and $\eta = 20\%$. Steeper switch-off leads to coherent absorption and, hence, a mixed-character memory, as seen by an increase in both $\tilde{\mathcal{C}}$ and $\eta$.
(b) $\tilde{\mathcal{C}}$ values (diamonds) and memory efficiencies (circles)  for different control profiles in (a).
}
\label{fig:eit_temporal_profile}
\end{figure}

Second, a non-optimized temporal profile has different effects in the narrow- and broadband EIT memory. 
As stated earlier,~improper timing and/or gradient of switch-off during storage gives rise to incoherent absorption loss in the narrowband regime. In contrast, for the broadband regime, a non-ideal control profile may lead to coherent absorption that contributes towards the storage through the mechanism of ATS memory, which may even \emph{increase} the memory efficiency. This effect is illustrated in Figs.~\ref{fig:eit_temporal_profile}(a,b) for broadband-EIT storage using different temporal profiles at fixed strength. Applying a control field with earlier and/or steeper switch-off can result in significantly larger memory efficiency than that which is achievable via the true EIT scheme.~The increased $\mathcal{C}$-values for such profiles confirm that the storage actually takes the character of the ATS scheme, which is inherently more optimal than the broadband-EIT scheme for the given optical depth and bandwidth conditions (Sec.~\ref{sec:adiabaticity}). Figures~\ref{fig:character} and \ref{fig:eit_temporal_profile} also highlight the smooth crossover between the EIT and ATS schemes, showing how sensitive the memory character is to the strength and profile of the control field employed for broadband signal storage.

In the ATS protocol, the control field mediates signal absorption via the ATS peaks by coupling the photonic coherence directly to the polarization mode, which then evolves into the spin-wave mode for storage. In general, a wide splitting, associated with a large $\Omega_{\rm C}$, leads to insufficient signal absorption due to a reduction in the effective optical depth~($\tilde{d}=d/2F$).  This results in a portion of the input signal being directly transmitted (or unabsorbed) through the medium, which in turn, does not contribute to the storage process (referred to as the ``transmission loss''). On the other hand, small $\Omega_{\rm C}$ decreases the rate at which the exchange of coherence occurs between the polarization and spin/photonic modes. This causes the memory operation to suffer from polarization decoherence during the writing and reading stages, which manifests as spontaneous-emission loss.

The control-field optimization in the ATS protocol ensures a trade-off between these two sources of loss via a fixed ``pulse-area'' operation in the storage and retrieval stages, given by $\mathcal{A_{\rm C}}= \int_0^{\tau_{\rm s}} \Omega_{\rm C}(t'){\rm dt'}= 2\pi$.  Besides exactly mitigating the losses during the writing and readout stages, the $2\pi$ pulse ensures that no coherence remains in the polarization mode immediately after these stages, thereby providing a complete transfer from photonic to spin mode, or vice-versa. Furthermore, the fixed pulse-area-based optimization reduces the profile and strength degrees into a single joint degree of freedom.  In conjunction with independence from optical depth, this feature provides great flexibility and simplicity for finding the optimal control field when compared to the EIT scheme.  For a given optical depth, using the ATS-protocol, it is possible to achieve optimal memory efficiency using different combinations of temporal profile and strength of control field, all fulfilling the 2$\pi$ pulse condition.  As an example, for a broadband Gaussian signal pulse, an optimal ATS memory can be implemented using an interrupted control profile with a Rabi frequency of $\Omega_{\rm C}/2\pi= B$ or using a control with the same spatiotemporal profile as the input, but with a peak Rabi frequency of $\Omega_{\rm C}/2\pi = B\sqrt{{\pi}/{2 \ln 2}} $. Such flexibility (up to a certain degree) is possible with the EIT-scheme only in the limit of large optical depths.

\subsubsection{Optimal operation: adiabatic vs.\ non-adiabatic (fast) memories} \label{sec:adiabaticity}

The distinct physical mechanisms underlying the EIT and ATS protocols lead to different requirements in terms of optical depth and bandwidth for optimal memory implementations.~In this section, we compare the optimality conditions of these protocols in both narrow- and broadband signal regimes for optimized control fields. Specifically, we seek an answer to the question: given a certain optical depth (or bandwidth), which bandwidths (or optical depths) are suited to reach the optimal memory efficiency using the EIT and ATS protocols?

\begin{figure*}[tb!]
    \centering
    \includegraphics[width = 17.7cm]{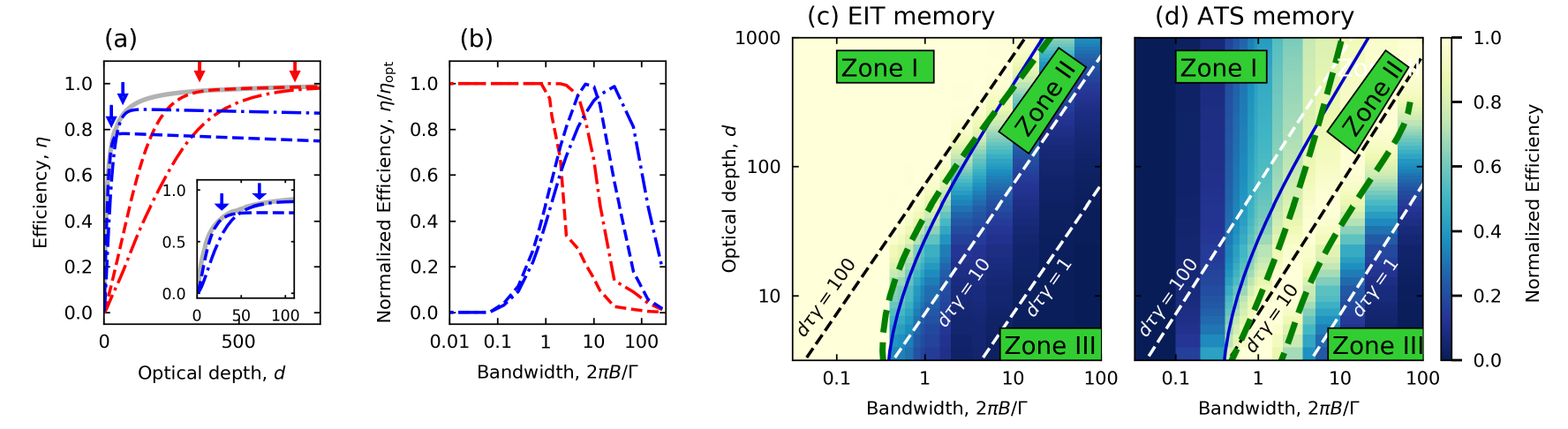}
    \caption{Adiabatic vs.\ non-adiabatic character in optimal EIT and ATS memories. (a)
    Efficiency  at fixed bandwidth [$ B = 5\Gamma/2\pi$ (dashed) and $10\Gamma/2\pi$ (dash-dot)] vs.\  optical depth using the EIT (red) and ATS (blue) protocols, along with the maximum achievable efficiency  (solid grey curve)~\cite{Gorshkov2007b} at each $d$.  The ATS-intersection (also shown in the inset) and EIT-merger points are indicated by arrows. (b) Optimality at fixed optical depth values \big[$d = {\rm 50}$ (dashed) and $\rm{250}$ (dash-dot)\big] vs.\ bandwidth. Efficiency is normalized with respect to the optimal efficiency for each optical depth. For moderate optical depths, the EIT curve (red) starts to deviate from optimality around $B \approx \Gamma / 2\pi$, with a rapid drop in efficiency thereafter. In the ATS-protocol, each optical depth is related to a unique bandwidth mode (in the broadband regime) for optimal storage. (c) Normalized efficiency (indicated by color) for varying bandwidths and optical depth in the EIT protocol depicting the full topology for optimal /  non-optimal operations. (d) As in (c), for the ATS protocol.}
    \label{fig:optimal_plots}
\end{figure*}

Figure~\ref{fig:optimal_plots}(a) shows the universal, protocol-independent optimal memory efficiency as a function of the optical depth (solid curve)~\cite{Gorshkov2007b}.   Numerically calculated efficiencies using the EIT and ATS protocols for narrow- and broadband signals are compared with the optimal efficiency values. In Fig.~\ref{fig:optimal_plots}(a),  points where the EIT or ATS memory curves coincide with the optimal memory curve correspond to configurations where optimality is satisfied. 

In the narrowband regime, the efficiency curves for EIT memory overlap entirely with the optimal memory curve, showing that optimal EIT memory can be implemented at any optical depth. On the other hand, the efficiency of narrowband ATS memory (not shown) stays near zero for all optical depths.~This is due to the fact that, in contrast to the broadband regime, the interaction time (given by the signal duration $\tau$) is longer that the coherence time ($1/\gamma$) between the ground and excited levels. Hence, coherence in the polarization mode, which is crucial to ATS-based storage, decays rapidly in proportion to $\exp(-\gamma\tau$).~This loss of coherence manifests as spontaneous emission. 

In the broadband regime, the EIT-efficiency curve does not entirely overlap with the optimal curve. Rather, it merges with the optimal curve only at high optical depths, and the merging point shifts toward greater optical depth values as the bandwidth is increased. In contrast, the efficiency curve for ATS memory never merges with the optimal memory curve, instead, it intersects the optimal curve only at certain specific optical depths. As the bandwidth is increased, the point of intersection moves toward higher optical depths. Importantly, the optical depth value at the ATS intersection point is about ten times smaller than the optical depth value at the EIT merging point for the same bandwidth.  This represents a key technical advantage of the ATS protocol for broadband signals.

A complementary analysis can be carried out for the memory efficiency as a function of signal bandwidth at fixed optical depths.  In this case, memory efficiency is normalized to the optimal efficiency at each optical depth, so that a value of unity for the normalized efficiency corresponds to configurations that satisfy optimality. As illustrated in Fig.~\ref{fig:optimal_plots}(b), a wide range of signal bandwidths (spanning the entire narrowband and into the beginning of broadband regime) is compatible with optimal EIT memory operation. However, only certain bandwidths in the broadband regime achieve optimal memory operation using the ATS protocol. Importantly for broadband signals, the largest bandwidth that can be optimally stored via the ATS protocol is nearly ten times larger than that achievable via the EIT protocol for the same optical depth. 

Finally, we combine the results of Figs.~\ref{fig:optimal_plots}(a,b) to show the full scope of optimal operation for EIT and ATS memories as a function of optical depth and bandwidth [Figs.~\ref{fig:optimal_plots}(c,d)]. In these figures, the region represented by ``Zone 1'' shows that the EIT protocol can be optimally implemented in the narrowband signal regime at any optical depth, and in the broadband signal regime only at very large optical depths. In this zone, the realization of optimal EIT relies upon satisfying the condition $d \gamma / B \gg 1$, which is a common feature of ``adiabatic quantum memories'' based on the elimination of absorption process as described in \cite{Gorshkov2007b}, including memories such as the the ''off-resonant Raman'' memory \cite{Nunn2007,Reim2010,Saunders2016}. On the other hand, the ATS protocol can be optimally implemented with only certain combinations of bandwidths and optical depths, corresponding to a narrow region in the broadband regime (``Zone 2''), where the EIT protocol is not optimal. In this region, the optimality of the ATS relies upon satisfying the condition $d \gamma /B =$ 8 to 10 (or equivalently $d / 2F = 3$ for backward retrieval). Furthermore, in the region indicated as ''Zone 3'', neither the ATS nor the EIT protocol can be fully optimal. However, in a significant portion of this zone, where the condition of $d \gamma /B =$ 1 to 8 is satisfied, the ATS protocol is much more efficient than the EIT protocol. In accordance with the universal classification of quantum memories, fulfillment of the condition $d \gamma /B \sim 1$, which characterizes efficient and optimal operation for ATS memories, is the common feature of ``non-adiabatic'' or ``fast'' memory protocols that are based on coherent absorption/re-emission processes such as the ``photon echo techniques''~\cite{Tittel2009}. The remaining portion of Zone 3, where  $d \gamma / B < 1$, is inaccessible to either protocol for efficient memory operation. Finally, we note that memory implementation for signal bandwidths that are on the order of transition linewidth ($B \approx \Gamma/2\pi$) corresponds to the crossover regime between both EIT and ATS \emph{phenomena} ($F \approx 1$) and optimal EIT and ATS \emph{memory operation}. 

These results conclusively show that the optimal operational regimes of the EIT and ATS protocols complement each other.~To realize an optimal memory using these schemes, the EIT protocol is the only option in the narrowband signal regime, while the ATS protocol is the best choice in the broadband regime, due to substantially less demand on optical depth and control-field power.

\section{EXPERIMENTAL DEMONSTRATIONS}
\label{sec:exp}

\begin{figure*}[tb!]
    \centering
    \includegraphics[width = 17.2 cm]{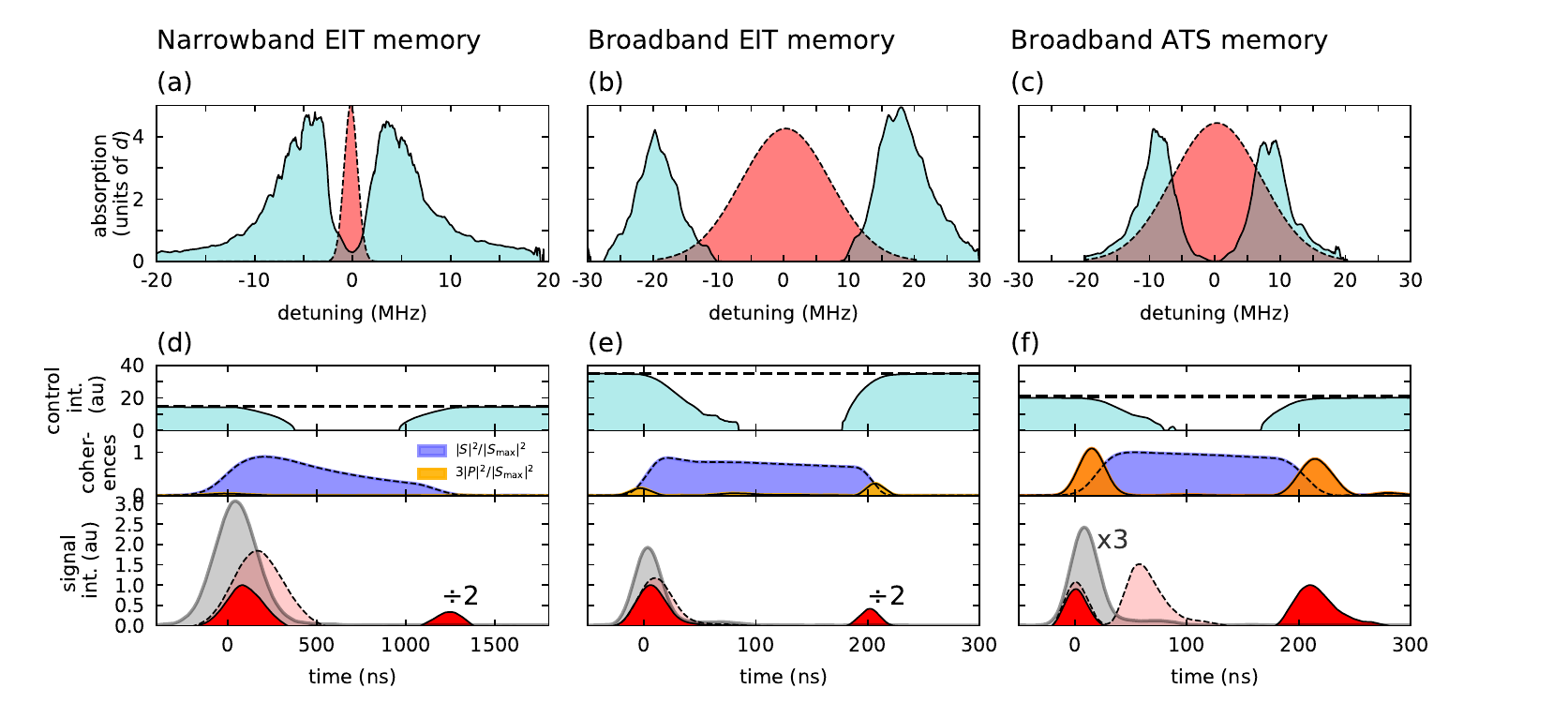}
\caption[Experimental demonstration]{Experimental demonstration of EIT and ATS memories in $^{\rm 87}$Rb atoms for narrowband [$B = 0.25\Gamma/2\pi$, (a) and (d)] and broadband [$B = 2.5\Gamma/2\pi$, (b), (c), (e)and (f)] signals. (a,b,c) Measured spectra of the absorption profile of the atomic medium (blue) and the input signal (red) for (a) $\Omega_{\rm C}/2\pi = 7.5~{\rm MHz}$, (b) $\Omega_{\rm C}/2\pi = 45~{\rm MHz}$, and (c) $\Omega_{\rm C}/2\pi = 15~{\rm MHz}$. (d,e,f) The time domain picture of signal delay and storage/retrieval processes associated with the top panels; coherence dynamics in the middle panels; and input (gray), delayed (light red), transmitted / leaked (red), on-demand recall (red) signals in the bottom panels. The coherences are shown at a given slice of the medium: $z = L/2$ for EIT memories and $z = 0$ for ATS memory. The control power and Rabi frequency are related as $\Omega_{\rm C}/2\pi = \alpha \sqrt{P}$, where $\alpha = {\rm 5.3~MHz/\sqrt{mW}}$.}
\label{fig:timefreqexp}
\end{figure*}

\begin{figure}
    \centering
\includegraphics[width =8.6 cm]{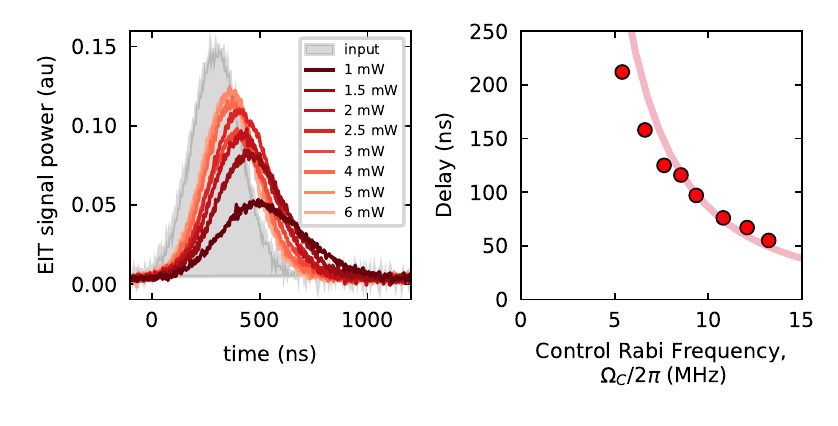}
\caption{Delay characterization for the narrowband EIT protocol.~The Gaussian input (shaded) has a bandwidth $B_{\rm narrow} = 1.62~{\rm MHz} \approx 0.25\Gamma/2\pi$ and the delay is measured as the difference between the peak positions of input and delayed pulses. (a) Delayed pulses for varying control powers. (b) Measured delay times vs.\ $\Omega_{\rm C}$ (circles) and calculated $\tau_{\rm D}^{\rm EIT}$ (curve , assuming $d  = 9$).}
    \label{fig:eit_delay}
\end{figure}

\begin{figure}
    \centering
\includegraphics[width =8.6 cm]{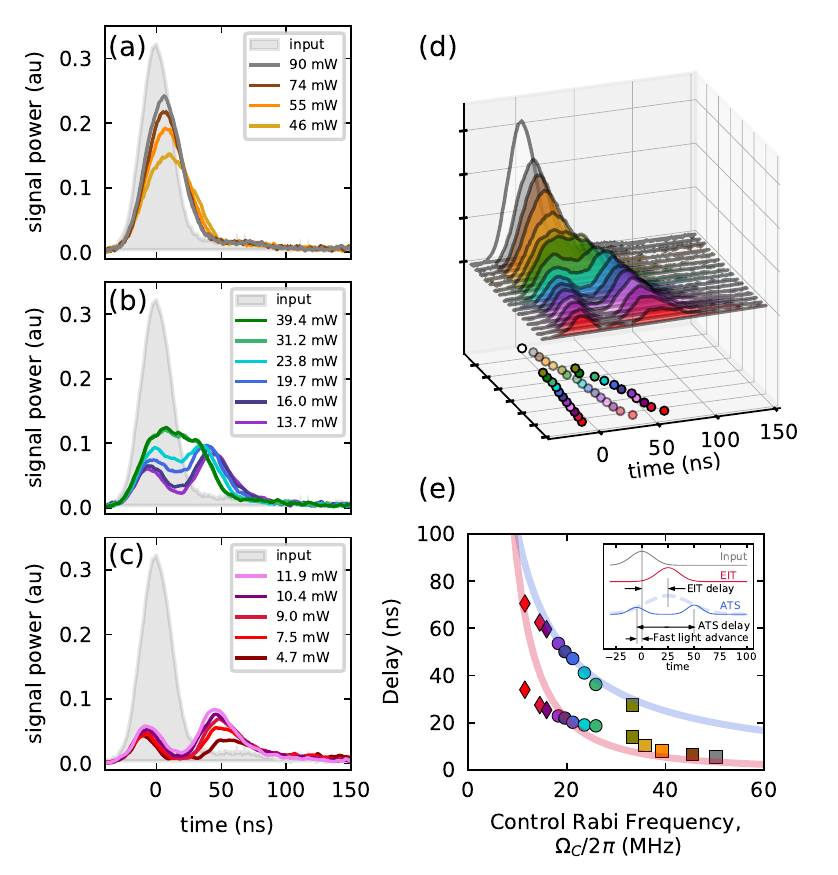}
\caption{Delay characterization for EIT, mixed, and ATS protocols in the broadband signal regime. (a) EIT-delay for control Rabi frequencies in the range $2.5B < \Omega_{\rm C}/2\pi < 4B$ (Here and in all parts, $\Omega_{\rm C}/2\pi = 5.3~{\rm MHz} \sqrt{P~[{\rm mW}]}$, where $P$ is indicated in legends.) (b) Mixed EIT-ATS delay for control Rabi frequencies $1.2B < \Omega_{\rm C}/2\pi < 2.2B$. (c) ATS delay for control Rabi frequencies $0.7B < \Omega_{\rm C}/2\pi < 1.3B$, where the ATS peaks' separation is close to the signal bandwidth. (d) Complilation of data in (a), (b), and (c) (same color coding), with the peak locations indicated in the lower plane: dark-filled circles indicate ATS peaks and light-filled are the EIT maxima. (e) Signal delay vs.\ control Rabi  frequency, with diamonds corresponding to EIT delays in (a), circles to mixed-character in (b), and squares to ATS delays in (c).  Inset shows delay definitions: ATS delay is the time between the transmitted and delayed peaks, and EIT delays are defined between the input signal peak and the output peak, with an estimate for the peak ``envelope'' made in the cases where ATS character is significant.}
    \label{fig:broadband_delay}
\end{figure}

In view of our theoretical analyses in the previous sections, we  present our experimental demonstrations, which highlight the essential distinctions between the EIT and ATS memory protocols. In these demonstrations, we implement each protocol using weak signal pulses (with a mean photon number that is much less than both control-field's typical photon number and the atom number) in a $\Lambda$-system [Fig.~\ref{fig:levels}(a)], that includes the D2 transitions from the hyperfine ground states of an ensemble of laser-cooled Rb-87 atoms (details of the experimental setup can be found in the supplementary information section of \cite{Saglamyurek2017}).  Following the methodology of our numerical analysis (Sec.~\ref{sec:Numerical Results}), we look into constant-control and on-demand memory implementations of the EIT and ATS protocols in both the narrow- and broadband signal regimes. Due to the ease of implementation, the delay and storage/recall operations are carried out in the forward retrieval scheme [Fig.~\ref{fig:levels}(b)], and our experimental parameters are as follows: the transition linewidth is $\Gamma = 2\gamma = 2\pi \times 6~{\rm MHz}$, the spin decoherence rate is $\Gamma_{\rm s} = 2\pi \times 0.12~{\rm MHz}$, the optical depth $d\approx 8~{\rm to}~10$, and the Rabi frequency $\Omega_{\rm C}$ ranges from $1.25\Gamma$ to $8\Gamma$. We note that for the optical depth in our setup, the maximum achievable forward mode efficiency is reasonably close to the optimal efficiency obtained via backward retrieval, and is thus, near-optimal. With these parameters and our technical limitations, we choose the signal durations to be $\tau =5/\gamma= 270~{\rm ns}$~($B \approx 0.25 \Gamma/2\pi$) for narrowband operation and $\tau = 0.5/\gamma = 27~{\rm ns}$~($B \approx 2.5\Gamma/2\pi$) for broadband operation, which can allow for near-optimal EIT and ATS memory, respectively.  

\subsection{Signal delay via EIT protocol vs.\ ATS protocol} \label{sec:delay_exp}

As detailed in Sec.~\ref{sec:disp vs abs}, slow-light induced delay in the EIT protocol and the absorption/re-emission mediated delay in the ATS protocol have distinct characteristics. This is a direct result of the different relationship between the power spectrum of the input signal and absorption spectrum of the atomic medium, which is determined by the strength of control-field. 

To delay a narrowband signal ($B_{\rm narrow} = 0.25 \Gamma/2\pi$) via the EIT protocol, we apply a control-field with constant $\Omega_{\rm C}/2\pi = 7.5~{\rm MHz}$ that is in the crossover between EIT and ATS regimes with $F \approx 1$, which induces a narrow transparency window of width  $\Delta \omega_{\rm EIT} =2\pi\times (3.2~${\rm MHz}) = $ 0.53 \Gamma$. In this configuration, the signal bandwidth lies inside the transparency window ($B < \Delta \omega_{\rm EIT}/2\pi < \Omega_{\rm C}/2\pi$) such that signal absorption is largely eliminated~[Fig.~\ref{fig:timefreqexp}(a)]. Concurrently, the steep dispersion associated with the EIT window leads to slow-light, resulting in a measured group delay of 130~ns [bottom panel in Fig.~\ref{fig:timefreqexp}(d)], which is in close agreement with the theoretically expected delay of 148~ns.  This yields a fractional delay $\tau_{\rm D}/ \tau \approx 0.5$, which although less than 2 is still sufficient for compressing (trapping) the signal inside the atomic medium for subsequent memory implementation with optimal efficiency, as detailed in the next section. 

To delay a broadband signal ($B_{\rm broad}=10 B_{\rm narrow}$) via the EIT protocol, a natural approach (with the given optical depth) would be to maintain the fractional delay from the narrowband EIT, by increasing the Rabi frequency by a factor of $\sqrt{\rm 10}$, which would broaden the transparency window and reduce the group delay by a factor of 10. However, since this strategy applies only in the narrowband regime, the scaling would result in the ATS peaks lying inside the expected transparency window ($B < \Omega_{\rm C}/2\pi < \Delta \omega_{\rm EIT}/2\pi$), leading to significant signal absorption. Therefore, an appropriate strategy would be to adjust the spacing between the Autler-Townes peaks to be broader than the signal bandwidth: $\Omega_{\rm C}/2\pi \approx 4B$. 

In our experiments with limited control power, we nearly satisfy this condition by setting $\Omega_{\rm C}/2\pi \approx 3B$ [Fig.~\ref{fig:timefreqexp}(b)]. This in turn corresponds to an increase in the control power by about 35 times compared to the narrowband case. Under this condition, the signal delay was measured to be 7.0~ns, yielding a fractional delay $\tau_{\rm D}/ \tau = 0.26$~[bottom panel in Fig.~\ref{fig:timefreqexp}(e)], which is significantly smaller than the one achieved for narrowband EIT. This is because at the expense of reducing the absorption for the true EIT operation, the slope of dispersion also got significantly decreased.  Consequently, insufficient spatial compression of this broadband signal makes the subsequent storage process non-optimal, as discussed in the next section. 

In contrast to the non-overlapping spectral conditions imposed by the EIT scheme, constant control delay of a broadband signal in the ATS protocol requires a significant overlap between the signal bandwidth and the ATS peaks. In our experiment, we satisfy this condition optimally by setting $\Omega_{\rm C}/2\pi = B_{\rm broad}$ [equivalently providing a pulse area of $\mathcal{A_{\rm C}} = 2\pi$, see Fig.~\ref{fig:timefreqexp}(c)], which amounts to about 9 times less control power than is required for broadband EIT. In this configuration, the resulting ATS delay is 57~ns [bottom panel in Fig.~\ref{fig:timefreqexp}(f)], which is in reasonably good agreement with the theoretically expected delay of $2\pi/\Omega_{\rm C} = 61~{\rm ns}$.  Here, the corresponding fractional delay $\tau_{\rm D}/ \tau \approx 2$, is fixed and significantly larger than the one achieved with our EIT-based implementations.  This important feature of the ATS-based delay makes signal trapping possible with almost no leakage loss, unlike the slow-light based fractional delay which depends on the optical depth. On the other hand, in place of signal leakage, the ATS protocol features transmission loss due to insufficient optical depth, emerging as the non-absorbed (directly transmitted) part of the signal with nearly zero delay [Fig.~\ref{fig:timefreqexp}(f)].

Additionally, we investigate the variation of delay time with respect to $\Omega_{\rm C}$ at fixed bandwidths for both the EIT and ATS protocols. In our narrowband EIT implementation, the group delay follows an inverse-squared dependence on $\Omega_{\rm C}$, which is in good agreement with the theoretically predicted $\tau_{\rm D}^{\rm EIT} = d\Gamma/\Omega_{\rm C}^{\rm 2}$, as shown in Fig.~\ref{fig:eit_delay}(b). Moreover, we observe that an increased group delay (with decreased $\Omega_{\rm C}$) is accompanied by a decrease in the delayed signal intensity [Fig.~\ref{fig:eit_delay}(a)]. This is because as $\Omega_{\rm C}$ is decreased, $\Delta \omega_{\rm EIT}$ also decreases, leading to more signal absorption, which results in loss via spontaneous emission. In the narrowband EIT protocol, there is a trade-off between this absorption loss and the leakage loss described above. 

Next, we characterize the ATS-based delay (predominantly in the ATS regime) by varying $\Omega_{\rm C}$ in a limited range that largely satisfies the bandwidth matching condition of the ATS protocol: $0.7B< \Omega_{\rm C}/2\pi < 1.3B$ [Fig.~\ref{fig:broadband_delay}(c)]. In this range, as in the EIT-delay, a lower $\Omega_{\rm C}$ leads to an increase in the delay time.~However, as also confirmed by our measurements, the ATS-delay bears an inverse-linear relationship to $\Omega_{\rm C}$, in contrast to the inverse-squared relation seen in the slow-light delay. Moreover, similar to the EIT-based delay, an increase in the delay time results in a decreased intensity of the delayed signal (eventually manifesting as spontaneous emission loss). However, signal loss in the ATS protocol is due to polarization decoherence in the ``desired'' absorption/re-emission processes, which is different from the loss mechanism of the EIT scheme that is mainly due to ``undesired'' incoherent absorption. 

To characterize the broadband EIT-delay (in the ATS regime), $\Omega_{\rm C}$ is varied in the range $\Omega_{\rm C}/2\pi > 2.25B$, where absorption is largely eliminated and delay is thus dominated by dispersion. In this configuration, as illustrated in Fig.~\ref{fig:broadband_delay}(a), the general characteristics and the $\Omega_{\rm C}$-dependence of signal delay are the same as in the narrowband EIT protocol (and hence quite different from the ATS protocol). However, an important distinction between the narrow- and broadband EIT-based delay emerges when comparing the role of ``undesired'' absorption in the scenarios where the signal spectrum partially overlaps with the absorption features: $B \approx \Delta \omega_{\rm EIT}/2\pi < \Omega_{\rm C}/2\pi$~(narrowband signals) and $B \approx \Omega_{\rm C}/2\pi$~(broadband signals). For the former case, as highlighted before, it is an incoherent process that acts a source of signal loss, while for the latter, it is a coherent process that can eventually lead to a transition to the ATS scheme, as investigated next.

To experimentally observe the transition between the signal delay mechanisms of the EIT and ATS schemes, $\Omega_{\rm C}$ is varied in the range $1.2B < \Omega_{\rm C}/2\pi < 2.5B$, as illustrated in Fig.~\ref{fig:broadband_delay}(b). For smaller bandwidths lying close to the ATS protocol's spectral matching condition, the transmitted and delayed parts of the signal are still distinct and separated in time by $2\pi/\Omega_{\rm C}$.~As $\Omega_{\rm C}$ is increased, these parts merge towards each other as the transmitted peak shifts forward in time and the delayed peak moves backward.  For these intermediate bandwidths, the amount of signal delay due to the slow-light and absorption/re-emission processes becomes comparable, and the system exhibits the characteristics of both delay mechanisms at the same time.  Upon further increasing $\Omega_{\rm C}$,~the transmitted and the ATS-delayed parts of the signal merge into a single pulse envelope, whose group delay is determined by the slow-light effect. This is the beginning of the regime where the true EIT scheme is in effect [Fig.~\ref{fig:broadband_delay}(a)].

Figures~\ref{fig:broadband_delay}(d,e) show the complete transition, as a function of control power, from the ATS- to mixed- to EIT- based delays, in the broadband regime.  

\subsection{On-demand memory implementation via EIT scheme vs ATS scheme} \label{sec:memory_exp} 
In light of our discussion on signal delay in the previous section, we now compare the features of on-demand memory implementation using the EIT and ATS protocols.  For these measurements, the control-field Rabi frequency was dynamically varied so as to generate an ``interrupted'' control profile.

For the narrowband EIT memory, we begin by optimizing the strength of the control field so as to realize the best possible signal compression (trapping) in the medium with minimal absorption loss (within the constraints of our setup).  As described in the previous section, we satisfy this trade-off with $\Omega_{\rm C}/2\pi \approx 8~\text{MHz}$, which leads to a fractional delay of 0.5 and 35$\%$ signal transmission. Next, we optimize the temporal profile of the control field, paying particular attention to the switch off/on phases where the coherent coupling between spin and photonic modes must be adiabatically maintained. We satisfy this condition by smoothly ramping down (up) the field to zero (maximum) to initiate storage (retrieval), as shown in the top panel in Fig.~\ref{fig:timefreqexp}(d).

With these control optimizations, we achieve a memory efficiency of 7.5$\%$ (including both storage and retrieval stages) for a storage time of 1.2~$\mu s$.  With the optical depth ($d \approx$ 10) and spin-decoherence rate corresponding to a memory decay time of $\tau_{\rm D} = 1/2\gamma_{\rm s} =$ 650~ns, the expected optimal memory efficiency for our setup is 8$\%$, which matches very well with the measured efficiency of our EIT memory implementation. This result is also consistent with the general adiabaticity condition for optimal EIT memory ($\tau d \gamma \gg 1$), which is fulfilled by $\tau d \gamma \approx 50$ in our experimental setting. Furthermore, using the experimentally established parameters, we numerically simulated our memory implementation, which is in close agreement with our experimental results. This analysis also allows us to establish the coherence dynamics [middle panel in Fig.~\ref{fig:timefreqexp}(d)], confirming that polarization coherence has not played a role in the storage and retrieval processes as expected for an adiabatic EIT memory. Finally, we calculate the memory character to be $\tilde{\mathcal{C}} = {\rm 0.06} \ll 1$, conclusively demonstrating the true EIT-character of this memory implementation. 

To implement the ATS memory, we ensure that the control pulse-area during writing and read-out stages is equal to 2$\pi$. For the constant control with $\Omega_{\rm C}/2\pi = B$ (as in Sec.~\ref{sec:delay_exp}), this condition is satisfied by abruptly switching-off the control field at $t = 2\pi/\Omega_{\rm C}$ (just before the pulse is about to begin re-emission) for storage, and then abruptly switching it back on after a desired storage time with the same control strength for retrieval [top panel in Fig.~\ref{fig:timefreqexp}(f)]. By implementing this operation within our technical limits, we achieve a memory efficiency of 23$\%$ for a storage time of 200~ns [bottom panel in Fig.~\ref{fig:timefreqexp}(f)], which is reasonably close to the expected optimal efficiency of 30$\%$ for our experimental conditions. This result is also in agreement with the general optimality condition for ATS memory ($\tau d \gamma \sim 1$), which is fulfilled by having $\tau d \gamma \approx 5$ in our setup.  Moreover, the numerical simulation of our implementation reveals that polarization coherence has played the essential role in the storage and retrieval processes [middle panel in Fig.~\ref{fig:timefreqexp}(f)], in contrast to the EIT scheme. These results, together with the memory character of $\tilde{\mathcal{C}} = 1$, demonstrate the true ATS character of this broadband memory implementation.

To demonstrate a broadband-EIT memory, the strength and profile of the control field are optimized by following the same general procedure as for our narrowband-EIT implementation. We achieve nearly optimized control-strength with $\Omega_{\rm C}/2\pi \approx 3B$ [amounting to roughly 9 times more power than demanded by the ATS memory, see top panels of Figs.~\ref{fig:timefreqexp}(e,f)] with a compromise between the requirements of minimal absorption and sufficiently large fractional delay (Sec.~\ref{sec:delay_exp}). Together with a sufficiently smooth switch-off/on for storage/retrieval, we realize broadband EIT memory with 10$\%$ efficiency for a 200~ns storage time [bottom panel in Fig.~\ref{fig:timefreqexp}(e)]. The memory efficiency in this case is significantly smaller than the ATS memory efficiency due to the fact that for the given combination of the optical depth and bandwidth, the adiabaticity condition cannot be satisfied for optimal EIT [i.e., EIT memory is inherently non-optimal in this regime, as indicated in Fig.~\ref{fig:optimal_plots}(c)].  Additionally, we calculate the memory character to be $\tilde{\mathcal{C}} = 0.2$, indicating that our implementation is predominantly characterized by the EIT scheme with some residual ATS contribution, which is also visible in the evolution of polarization coherences in Fig.~\ref{fig:timefreqexp}(e) (middle panel).  For a final verification, we numerically optimize the control strength further towards the true EIT memory.  In this situation, while obtaining the memory character of $\tilde{\mathcal{C}} < 0.1$, we estimate the memory efficiency to be $\approx 7\%$, which is smaller than the measured one, suggesting that in contrast to narrowband-EIT memory, the absorption of the signal has coherently contributed to this storage process via the mechanism of the ATS protocol.

\begin{figure}[t]
    \centering
    \includegraphics[width = 8.6 cm]{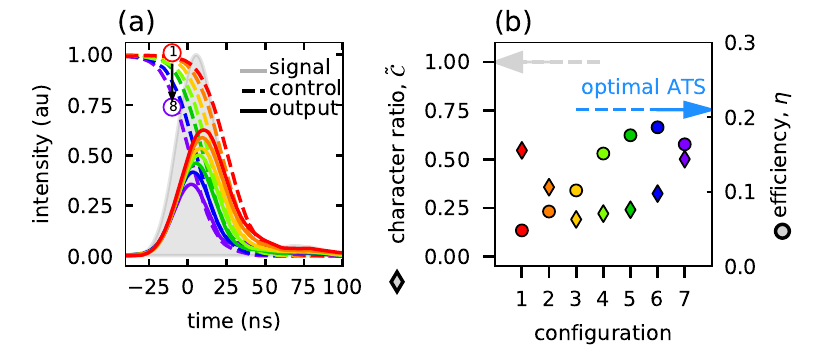}
    \caption{Experimental demonstration of the transition from EIT- to ATS-based broadband storage by varying the control-temporal-profile. Here $B = 16.3~{\rm MHz}$ and  $\Omega_{\rm C}/2\pi = 44~{\rm MHz}$. (a) Writing stage showing control fields at different switch-off (ramp down) positions. The leaked/transmitted signals correspond to the EIT/ATS dominated storage. (b) $\tilde{\mathcal{C}}$ values and the memory efficiencies $\eta$ for different control profiles. ``Configuration 3'' bears the maximum EIT-character with minimal value of $\tilde{\mathcal{C}} = 0.19$ corresponding to $\eta = 10\%$. While ``configuration 6'' gives the highest efficiency of 19$\%$, it benefits from the ATS storage ($\tilde{\mathcal{C}} = 0.35$), and the efficiency remains below the measured efficiency of 23$\%$ (blue arrow) for optimal ATS operation.}
    \label{fig:mixed_regime}
\end{figure}

Finally, we systematically demonstrate that implementing a broadband-EIT memory with a non-optimized control field can easily lead to a memory that operates, at least in part via the ATS protocol's mechanism. In this demonstration, we lay emphasis on the impact of the switch-off timing of the control field (during the writing stage), while keeping fixed the control strength and storage time used for the broadband-EIT implementation. After determining the optimal timing for (nearly) true-EIT operation [control profile in Fig.~\ref{fig:timefreqexp}(e)], we shift the moment in time  at which the control field begins to ramp down, both forward and backward, relative to the true-EIT configuration~[Fig.~\ref{fig:mixed_regime}(a)]. For each timing configuration, we measure the memory efficiency and estimate the memory-character parameter Fig.~\ref{fig:mixed_regime}(b)]. Our results show that certain configurations with non-optimized timing can yield memory efficiencies that are significantly larger than those expected from  true-EIT memories for broadband storage. In these cases, we find that the memory character $\tilde{\mathcal{C}} > 0.2$, which suggests that the memory operation has a mixed character where both EIT and ATS storage mechanisms play a role.~Efficiency-wise, this mixed operation greatly benefits from the inherent optimality of the ATS protocol, thereby outperforming the true-EIT protocol, which is non-optimal in our experimental regime. Nevertheless, the maximum efficiency is still lower than the efficiency of the control-optimized ATS memory. This suggests that direct implementation of the true ATS memory would be the best strategy, not only for reaching an optimal broadband memory with minimal demand, but also for eliminating the technical difficulty of truly isolating the EIT memory from having ATS character.

These results, together with those of Sec.~\ref{sec:delay_exp} that showed that a non-optimized control-field strength can also lead to a dual memory character, demonstrate that the transition between the EIT and the ATS memory protocols is  smooth.  This implies that in an experimental setting, a turn of the control-field knob in the pursuit of empirically optimized efficiency could readily lead to an inadvertent transition from one kind of memory to the other. If one is not aware of the recently proposed ATS protocol~\cite{Saglamyurek2017}, this situation can easily lead to misinterpreted memory implementations, particularly by an experimentalist interested in efficient storage of broadband photonic signals who uses the EIT protocol at limited optical depths. Indeed, this raises the question of whether some previously reported broadband quantum memories based on the EIT protocol~\cite{Ding2013,Gouraud2015} may have benefited in-part or in-whole from the ATS mechanism, and if those systems would further benefit from using the true-ATS protocol.

\section{CONCLUSION}
\label{sec:conclude}
In this work, we  comprehensively analysed  the distinctions between  optical memory implementations using the EIT and ATS protocols, and supplemented these results with experimental demonstrations. We emphasize throughout that the EIT and ATS memory protocols -- associated with the ways in which a quantum memory is operated and responds -- are distinct from the EIT and ATS regimes of coupling strength. Indeed, we show that the EIT memory protocol can be implemented in the ATS regime of coupling strength, albeit with high demands on technical resources. 

The differences between the EIT and ATS protocols arise from their origins at the opposite limits of light-matter interaction -- adiabatic elimination of signal absorption vs.\ mediation by signal absorption --  and dictate the optimal memory conditions for each protocol. The EIT protocol is favorably implemented in the EIT regime and offers the best choice for optimally storing narrowband signals, whereas the ATS protocol is intrinsically suited for implementation in the ATS regime for optimally storing broadband signals. In this way, the optimal efficiency landscape of these protocols complement each other in a remarkable way.

This work also includes a detailed investigation of broadband signal storage via the EIT protocol, which has been largely overlooked. Although the storage mechanism for both narrow- and broadband EIT memories is the same, an important difference comes from the role of undesired absorption. It is well-known that any residual absorption appears as an incoherent loss in the narrowband EIT memory. However, as explored in this study, the same process in the broadband-EIT protocol leads to useful coherent storage via the mechanism of the ATS protocol. 

Finally, we explored at those cases in the broadband signal regime where the storage exhibits a mixed character of both the EIT and ATS protocols. We showed that the transition from the true-EIT memory to the one having mixed-character storage is gradual and can occur by simple variation of the strength and/or temporal profile of the control field. Moreover, to estimate the dominant storage mechanism for a given configuration, we have defined a useful, dimensionless parameter called the memory character factor $\mathcal{C}$ and have established bounds on it for the true EIT, true ATS and mixed storage operations. Specifically, we have shown that  implementing an efficient EIT memory that avoids any ATS character requires significant demands on control power and optical depth when compared to the true-ATS storage. This agrees well with the optimality conditions determined here and in other work.

Our investigation adds a fresh applications-based perspective to the ongoing discussion centered on phenomenon-based comparisons between EIT and ATS~\cite{Abi-Salloum2010,Anisimov2011a,Giner2013,Sun2014,Lu2015,Tan2014,Hao2017}. Further, it clarifies any potential confusion that may arise between the two perceivably similar EIT and ATS memory protocols, which is important because by understanding the underlying mechanisms of these memories, a proper approach to optimization can be taken: one that  yields the best results for a given set of technical and design limitations.  Though we have concentrated on the atomic three-level system in this work, we expect that these results are broadly applicable to other atom-like, spin-wave quantum memory media, including colour centres~\cite{Siyushev2017,Zhou2017}, optomechanical systems~\cite{Agarwal2009,Weis2010,Teufel2011a,Safavi-Naeini2011,Lake2018}, superconducting qubits~\cite{Sillanpaa2009,Abdumalikov2010a,Leung2012,Novikov2013,Sun2014}, and rare-earth systems~\cite{Afzelius2010,Gundoan2015}.

\acknowledgments{We thank Khabat Heshami for ongoing discussions and for sharing code for the initial Maxwell-Bloch simulations. We appreciate generous technical support from Greg Popowich and the following groups for lending us equipment for our initial measurements: J.\ Beamish, J.\ P.\ Davis,  F.\ Hegmann, A.\ Lvovsky, W.\ Tittel, R.\ Wolkow. We also thank Dr.\ Barry Sanders, Dr.\ Ying-Cheng Chen and Dr.\ Chris O'Brien for useful discussions. We gratefully acknowledge funding from the Natural Science and Engineering Research Council of Canada (NSERC RGPIN-2014-06618), Canada Foundation for Innovation (CFI), Canada Research Chairs Program (CRC), Canadian Institute for Advanced Research (CIFAR), Alberta Innovates (AITF), and the University of Alberta.}



\end{document}